\documentclass[twoside,12pt, a4paper, footinclude=true, headinclude=true, cleardoublepage=empty, xcolor=table]{scrbook}
\usepackage[linedheaders,parts,pdfspacing,dottedtoc]{classicthesis}
\usepackage{amsmath}
\usepackage{acronym}
\usepackage[a4paper,width=150mm,top=25mm,bottom=25mm]{geometry}
\usepackage{eso-pic} 
\usepackage{graphicx} 
\graphicspath{ {Images/} }
\usepackage[lofdepth,lotdepth]{subfig}
\usepackage[utf8]{inputenc}
\usepackage[english]{babel}
\usepackage{csquotes}
\usepackage{bookmark}
\usepackage{longtable}
\usepackage{array}
\usepackage{multirow}
\usepackage{times}
\usepackage{lingmacros}
\usepackage{color, colortbl}
\usepackage{tabularx}
\usepackage{pdfpages}
\usepackage{footnote}
\usepackage{tikz}
\usepackage{mathdots}
\usepackage{yhmath}
\usepackage{cancel}
\usepackage{siunitx}
\usepackage{amssymb}
\usepackage{gensymb}
\usepackage{booktabs}
\usepackage{import}
\usetikzlibrary{fadings}
\usetikzlibrary{patterns}
\definecolor{mygray}{rgb}{0.86,0.86,0.86}
\makesavenoteenv{tabular}
\makesavenoteenv{table}
\usepackage{multirow}

\newcommand{\bs}[1]{\boldsymbol{#1}}

\def\signed #1{{\leavevmode\unskip\nobreak\hfil\penalty50\hskip2em
  \hbox{}\nobreak\hfil(#1)%
    \parfillskip=0pt \finalhyphendemerits=0 \endgraf}}

    \newsavebox\mybox

\usepackage[square]{natbib}
\def \ColourPDF {Images/nat-farve.pdf}
\def \TitlePDF   {Images/ku-en.pdf}

\subject{ 
  \vspace{3.5cm}
  \small{Faculty of Science} \\
  \Large{Master's Thesis} \\}

\title{
  User-click Modelling for Predicting Purchase Intent
}

\author{
  \Large{Simone Borg Bruun} \\
  \texttt{clx749@alumni.ku.dk}
  \vspace{1.5cm} \\
  \large{Supervisors: Christina Lioma, Maria Maistro, Casper Hansen, Christian Hansen }
}
\date{March 2020}

\begin{document} 
\pagenumbering{roman}
\AddToShipoutPicture*{\put(0,0){\includegraphics*[viewport=0 0 700 600]{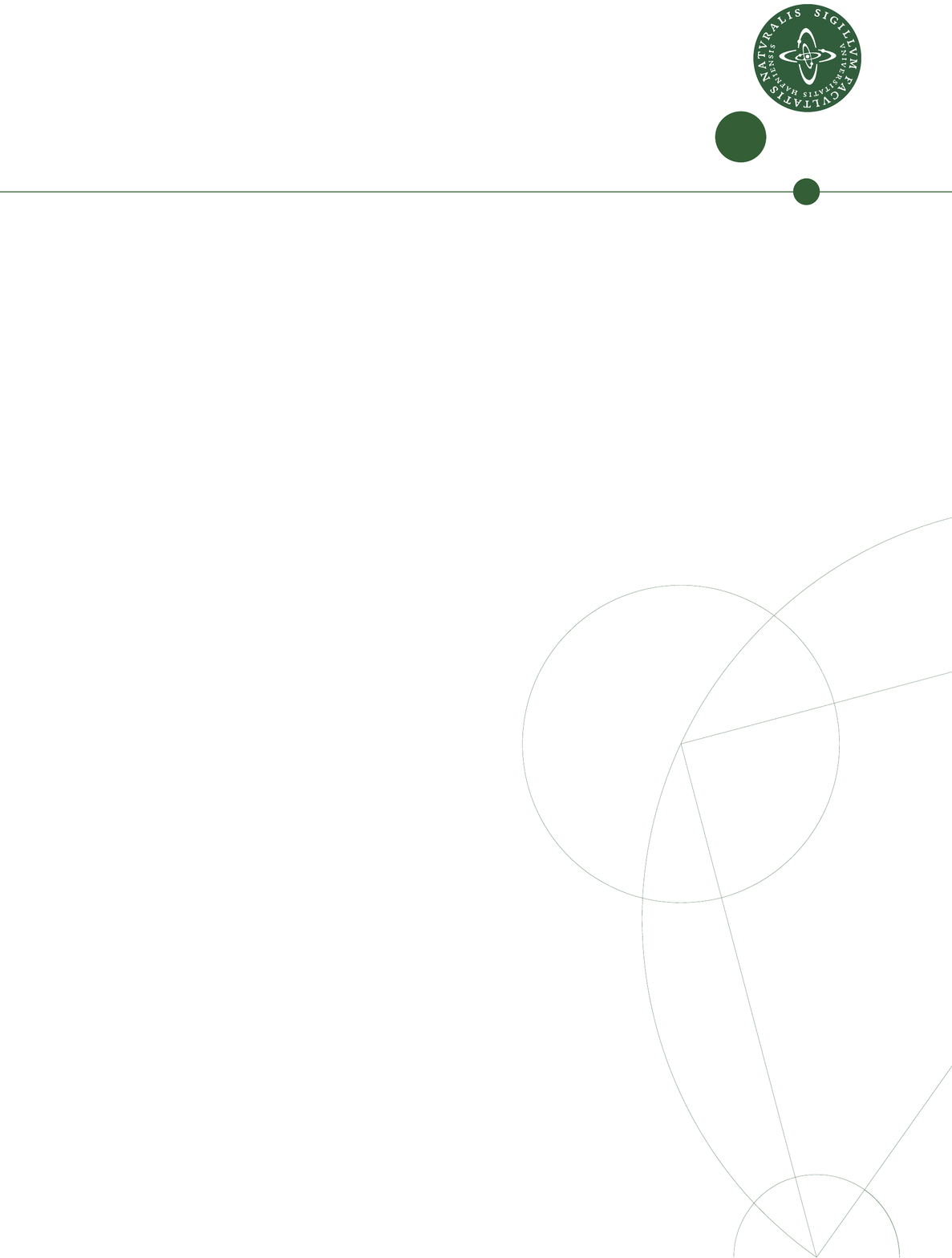}}}
\AddToShipoutPicture*{\put(0,602){\includegraphics*[viewport=0 600 700 1600]{\ColourPDF}}}
\AddToShipoutPicture*{\put(0,0){\includegraphics*{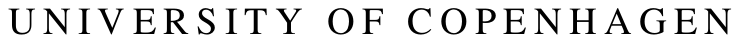}}}

\clearpage\maketitle
\thispagestyle{empty}

\newpage

\pdfbookmark[1]{Abstract}{Abstract}
\chapter*{Abstract}
This thesis contributes a structured inquiry into the open actuarial mathematics problem of modelling user behaviour using machine learning methods, in order to predict purchase intent of non-life insurance products.\\
It is valuable for a company to understand user interactions with their website as it provides rich and individualized insight into consumer behaviour. Most of existing research in user behaviour modelling aims to explain or predict clicks on a search engine result page or to estimate click-through rate in sponsored search. These models are based on concepts about users’ examination patterns of a web page and the web page’s representation of items.\\
Investigating the problem of modelling user behaviour to predict purchase intent on a business website, we observe that a user’s intention yields high dependency on how the user navigates the website in terms of how many different web pages the user visited, what kind of web pages the user interacted with, and how much time the user spent on each web page. Inspired by these findings, we propose two different ways of representing features of a user session leading to two models for user click-based purchase prediction: one based on a Feed Forward Neural Network, and another based on a Recurrent Neural Network.\\
We examine the discriminativeness of user-clicks for predicting purchase intent by comparing the above two models with a model using demographic features of the user. Our experimental results show that our click-based models significantly outperform the demographic model, in terms of standard classification evaluation metrics, and that a model based on a sequential representation of user clicks yields slightly greater performance than a model based on feature engineering of clicks.

\pdfbookmark[1]{\contentsname}{tableofcontents}

\setcounter{tocdepth}{2} 
\setcounter{secnumdepth}{3} 

\tableofcontents 




\pagenumbering{arabic}

\chapter{Introduction}\label{chap:introduction}
This thesis presents a structured inquiry into the open actuarial mathematics problem of modelling user behaviour using machine learning methods, in order to predict purchase intent of non-life insurance products: The objective of the thesis is to model user behaviour on a service-based business website carrying out a predictive modelling approach. The model will operate on click log data collected from the website of an insurance company. In addition, we will investigate the ability of click data to predict purchase intent, compared to demographic data describing characteristics such as age, gender and income.
\\\\
Among the huge number of websites on the world wide web, there exist several types of websites trying to accomplish different goals. By now, almost every business has a website, which communicates the types of products or services the business offers.
Understanding user behaviour on a business website can be valuable for the company, as such insights can be used to improve the website to better serve users. A way to obtain insights about user behaviour is through data collected from the website, and the challenge is to break that data into valuable knowledge. Moreover, the website data is a source of knowledge about potential customers, as well as additional knowledge about existing customers. Thus it can be further valuable for the company, if this data can help identifying aspects of customer behaviour such as consumer preferences and intentions.\\
An insurance company is a service-based business providing intangible products. Like many other companies, insurance companies also need to follow the digitization by increasingly offering services online. Insurance companies are already highly data-driven companies, so it is particularly befitting for them to further gain access of website data. 
\\
In non-life insurance, Generalized Linear Models (GLM) are very popular for modelling risk premium as they can handle regression in for example the Poisson and Gamma families, which are useful to model claim numbers and claim sizes, respectively. But when it comes to other modelling problems in non-life insurance, such as customer churn or insurance fraud, GLM may not be the most suitable framework, leading to a need of exploring more appropriate statistical methods. This is the case for user behaviour as well. 
\\
In the field of customer behaviour analysis, demographic data is widely used to segment customers into groups with shared characteristics. Demographic data includes general information about individuals such as age, gender, ethnicity, type of employment, education, marital status, and so on. Demographic segmentation assumes that customers with similar demographic profiles will exhibit similar motivations, interests and lifestyles, and that these characteristics will translate into similar preferences and behavioural patterns.
\\\\
In the remainder of the thesis, chapter \ref{chap:relatedwork} presents related work in the area of click models and discusses strengths and weaknesses of the state of the art. In chapter \ref{chap:data} we introduce the data set to be used and analyze the data in relation to the predicting task. In chapter \ref{chap:models} we propose two different click-based models that are aimed at modelling user behaviour on an entire website as well as predicting purchase intent. In chapter \ref{chap:experiment} we evaluate and investigate our proposed models. The thesis ends with a conclusion and a discussion of future work in chapter \ref{chap:conclusion}.

\chapter{Related Work}\label{chap:relatedwork}
\pdfbookmark[1]{Basic Click Models}{basic}
\section{Basic Click Models}\label{sec:basic}
Users commonly interact with the web through clicks. Based on collected click log data, a click model aims to model web user behaviour often with the purpose of predicting clicks. A great deal of previous work on click modelling concerns web search. Such models describe user behaviour on a search engine result page (SERP). A search engine returns a listing of items in response to a user's information need, requested by a query. The items can be web pages or documents, often presented with snippets and identified by their URLs. In this case the click log data set typically consists of user IDs with issued queries, query timestamps, item clicks and rank of items. \\
The basic click models in this area are based on the probabilistic graphical model (PGM) framework \citep{KollerFriedman2009}. In this framework, user search behaviour is treated as a sequence of observable and hidden events. Letting $u$ denote an item, the event of a click can be represented by the binary random variable $C_u$. Besides the click event, most probabilistic click models consider two more events: a user examines an item, and a user is attracted by the item. These events are represented by the binary random variables $E_u$ and $A_u$, respectively. Furthermore, most models include a so-called \textit{examination hypothesis}:
\begin{align}
C_u = 1 \Leftrightarrow E_u = 1 ~\text{and}~ A_u = 1,
\end{align}
which means that a user clicks on an item if, and only if, the user examined the item and was attracted by it. The random variables $E_u$ and $A_u$ are usually assumed independent. The attractiveness probability is usually modeled with a parameter that depends on the query-item pair. The examination probability is modeled differently by different click models, with the majority assuming the \textit{position bias}: search engine users tend to click more frequently on items higher in the ranking. \\
The \textit{position-based model} (PBM) \citep{ChuklinAleksandr2015} introduces an examination parameter that depends on the rank $r$ of the item. Thus the examination probability decreases as the user goes down the SERP. The model can be written as follows:
\begin{align}
P(C_u = 1) &= P(E_u = 1) \cdot P(A_u = 1) \\
P(A_u = 1 ) &= \alpha_{uq} \\
P(E_u = 1) &= \gamma_{r},
\end{align}
where $\alpha_{uq}$ and $\gamma_{r}$ denote parameters, whose values typically range uniform.\\
The \textit{cascade model} (CM) \citep{ChuklinAleksandr2015} assumes that a user scans a SERP from top to bottom until the user finds a relevant item. Under this assumption the model can be formalized as follows:
\begin{align}
P(C_{u_r} = 1) &= P(E_{u_r} = 1) \cdot P(A_{u_r} = 1) \\
P(A_{u_r} = 1 ) &= \alpha_{uq} \\
P(E_{u_1} = 1) &= 1 \label{first item}\\
P(E_{u_r} = 1 | E_{u_{r-1}} = 0) &= 0 \label{next items 1}\\
P(E_{u_r} = 1 | C_{u_{r-1}} = 1) &= 0 \label{next items 2}\\
P(E_{u_r} = 1 | E_{u_{r-1}}=1, C_{u_{r-1}}  = 0) &=1 \label{next items 3},
\end{align}
where (\ref{first item}) means that the user always examines the first item, while (\ref{next items 1}), (\ref{next items 2}) and (\ref{next items 3}) mean that items at bigger ranks are examined if, and only if, the previous item was examined and not clicked.
Hence the CM implies that all items up to the first-clicked item were examined, and a user who clicks never comes back. This limits its applicability to sessions with one click. This problem has been addressed in the \textit{user browsing model} (UBM), the \textit{dependent click model} (DCM), the \textit{click chain model} (CCM) and the \textit{dynamic Bayesian network} model (DBN) \citep{ChuklinAleksandr2015}, which are all extensions of the CM.
In UBM the examination parameter does not only depend on the rank of an item but also on the rank of the previously clicked item. The DCM, CCM and DBN introduce different continuation parameters to handle sessions with multiple clicks and model user satisfaction.\\
These models are powerful as they allow one to express many ideas about user behaviour and incorporate additional signals. Although, these models are limited by the specification of simple dependencies between user's behavioural states, which might be more complex.


\pdfbookmark[1]{Advanced Click Models}{advanced}
\section{Advanced Click Models}\label{sec:advanced}
A wide range of click models that somehow improve over the basic click models have been proposed. That includes extensions of the basic click models for web search and modification of the models to handle other click modelling areas. 
\\\\
The basic click models presented above model user behaviour based on individual queries. However, user sessions might include multiple queries with dynamic interactions. \citet{ZhangYuchen2011} present a general notion of a search task proposing a \textit{task-centric click model} (TCM). They consider the sequence of queries and their subsequent clicks in a search session as representing the same task from the user's perspective. Based on a qualitative analysis Zhang et al. formalize two new assumptions as the basis for user modelling. The first one is the \textit{query bias}: if a query does not match the user's information need, the query is reformulated to better match the information need. The second one is the \textit{duplicate bias}: if an item has been examined before, it has a lower probability of being clicked when the user examines it again. \\
Letting $i$ indicate the $i$'th query and $i'$ the latest query where an item has appeared, the TCM is formalized as follows:
\begin{align}
P(C_{u_r} = 1) = P(M_i = 1) &\cdot P(F_{i,{u_r}} = 1) \cdot P(E_{i,{u_r}} = 1) \cdot P(A_{i,{u_r}} =1) \\
P(M_i = 1 ) &= \beta_1 \\
P(N_i = 1 | M_i = 0) &= 1 \\
P(N_i = 1 | M_i = 1 ) &= \beta_2 \\
P(H_{i,{u_r}} =1 | H_{i',{u_{r'}}} = 0, E_{i',{u_{r'}}}=0) &= 0 \\
P(F_{i,{u_r}} = 1 | H_{i,{u_r}} = 0) &= 1 \\
P(F_{i,{u_r}} = 1 | H_{i,{u_r}} = 1 ) &= \beta_3 \\
P(E_{i,{u_r}} = 1 ) &= \gamma_{r} \\
P(A_{i,{u_r}} = 1) &= \alpha_{uq}  ~~,
\end{align}
where $M_i$ indicates whether the $i$'th query matches the user's intent, $N_i$ indicates whether the user submits another query, $H_{i,{u_r}}$ is previous examination of the item, and $F_{i,{u_r}}$ is freshness of the item. \\
The TCM can be built on top of any basic click model, which determines the attractiveness and examination parameters.
\\\\
Many recent studies showed that there is a large proportion of non-linear examination and click behaviour in web search, which the basic click models, including TCM above, fail to cope with. Moreover, the examination hypothesis is conflicted when modelling other search behaviour such as aggregated search\footnote{Aggregated search is a search paradigm where a SERP aggregates results from multiple sources known as verticals (e.g. News, Image or Video verticals).}, due to the representation of the items. For example, vertical blocks on an aggregated SERP can be more visually salient and attract more user attention than other results on the SERP.\\
The \textit{whole page click model} (WPC) by \citet{Chen2011} aims to tackle the problem of side elements on a SERP, such as ads or related searches. The model is designed as a nested structure with an outer layer to learn transition probabilities between blocks (organic results, top ads, side ads, related searches, etc.) and an inner layer to model user behaviour within a block. The outer layer is modeled via a Markov chain, while the inner layer is based on a basic click model. By assuming that each click solely depends on its current block, the transition probabilities are estimated by maximizing the following likelihood function:
\begin{align}
P(R_{\pi(s)},C) = \sum_{R_s \supset R_{\pi(s)}} \prod_i P(C_i|B_i)P(R_s),
\end{align}
where $R_s$ describes the sequence of blocks ($B_i$) that a user examines in session $s$, $R_{\pi(s)}$ describes a subsequence of $R_s$ containing blocks with clicks, and $C_i$ describes clicks in the $i$'th block.
\\\\
Overall, the PGM framework yields a mathematically suitable way to infer information given a set of events. However, one limitation is that the framework relies on a manual setting of dependencies between events. Different click models based on the PGM framework use different hand-crafted sets of dependencies. Both the basic click models and the advanced click models are, by necessity, simplifications and likely miss key aspects of user behaviour. Furthermore, they do not easily generalize to other click modelling areas.

\pdfbookmark[1]{Feature-based Click Models}{feature}
\section{Feature-based Click Models}\label{sec:feature}
Also feature-based approaches to predict clicks have been investigated. In the setting of sponsored search \cite{Richardson2007} suggest to use a feature-based logistic model to predict click-through rate (CTR) \footnote{CTR is the ratio of users who click on a specific item to the number of total users who viewed the item.}. Sponsored search is a service that displays advertisements (ads) on a SERP along with organic search results. The ads are targeting to the search query, and click predictions in sponsored search can have an immense influence on revenue.
They extract the features from the query, item and rank, and cope with the problem of predicting CTR of new ads.\\
Like logistic regression, decision tree is a general feature-based model that maps the relationship between features and a target. Decision trees can efficiently capture interactions between various features. \cite{Zengin2017} supplement the current query and item features with previous query and item features extracted from the user's session history, and use a tree-based model to predict a user click on a SERP. 
They show that including features from the session history improves the model performance. \\
Regression and tree-based models are popular models as it is easy to understand and interpret the model parameters. However, they can only work with tabular data, so in the case of sequence data an investment in feature engineering is required. 

\pdfbookmark[1]{Neural Click Models}{neural}
\section{Neural Click Models}\label{sec:neural}
\begin{figure}[]
\centering
\tikzset{every picture/.style={line width=0.75pt}} 

\begin{tikzpicture}[x=0.75pt,y=0.75pt,yscale=-1,xscale=1]

\draw   (150,120) .. controls (150,103.43) and (163.43,90) .. (180,90) .. controls (196.57,90) and (210,103.43) .. (210,120) .. controls (210,136.57) and (196.57,150) .. (180,150) .. controls (163.43,150) and (150,136.57) .. (150,120) -- cycle ;
\draw   (155,205) .. controls (155,191.19) and (166.19,180) .. (180,180) .. controls (193.81,180) and (205,191.19) .. (205,205) .. controls (205,218.81) and (193.81,230) .. (180,230) .. controls (166.19,230) and (155,218.81) .. (155,205) -- cycle ;
\draw    (180,180) -- (180,152) ;
\draw [shift={(180,150)}, rotate = 450] [fill={rgb, 255:red, 0; green, 0; blue, 0 }  ][line width=0.75]  [draw opacity=0] (8.93,-4.29) -- (0,0) -- (8.93,4.29) -- cycle    ;

\draw   (155,35) .. controls (155,21.19) and (166.19,10) .. (180,10) .. controls (193.81,10) and (205,21.19) .. (205,35) .. controls (205,48.81) and (193.81,60) .. (180,60) .. controls (166.19,60) and (155,48.81) .. (155,35) -- cycle ;
\draw    (180,90) -- (180,62) ;
\draw [shift={(180,60)}, rotate = 450] [fill={rgb, 255:red, 0; green, 0; blue, 0 }  ][line width=0.75]  [draw opacity=0] (8.93,-4.29) -- (0,0) -- (8.93,4.29) -- cycle    ;

\draw    (210,120) -- (240,120) ;

\draw    (240,80) -- (240,120) ;

\draw    (120,80) -- (240,80) ;

\draw    (120,80) -- (120,120) ;

\draw    (120,120) -- (148,120) ;
\draw [shift={(150,120)}, rotate = 180] [fill={rgb, 255:red, 0; green, 0; blue, 0 }  ][line width=0.75]  [draw opacity=0] (8.93,-4.29) -- (0,0) -- (8.93,4.29) -- cycle    ;

\draw (180,205) node [scale=0.9] [align=left] {$\displaystyle \boldsymbol{x}$};
\draw (180,120) node [scale=1] [align=left] {$\displaystyle \boldsymbol{h}$};
\draw (180,35) node [scale=0.9] [align=left] {$\displaystyle \boldsymbol{o}$};

\end{tikzpicture}
\caption{Folded diagram of an RNN with a loop.}
\label{fig:FOLDED}
\end{figure}
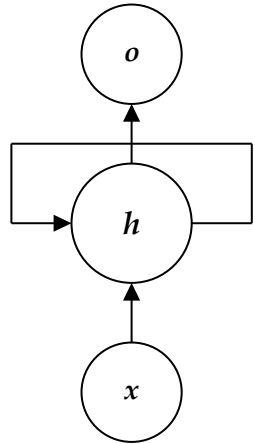

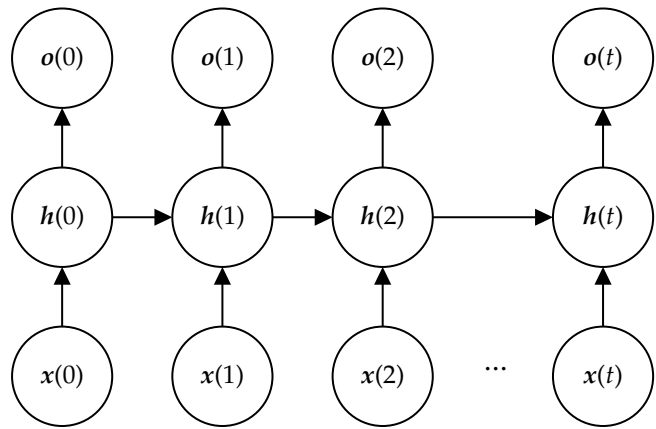
\begin{figure}[]
\centering
\tikzset{every picture/.style={line width=0.75pt}} 

\begin{tikzpicture}[x=0.75pt,y=0.75pt,yscale=-1,xscale=1]

\draw   (70,125) .. controls (70,111.19) and (81.19,100) .. (95,100) .. controls (108.81,100) and (120,111.19) .. (120,125) .. controls (120,138.81) and (108.81,150) .. (95,150) .. controls (81.19,150) and (70,138.81) .. (70,125) -- cycle ;
\draw    (120,125) -- (148,125) ;
\draw [shift={(150,125)}, rotate = 180] [fill={rgb, 255:red, 0; green, 0; blue, 0 }  ][line width=0.75]  [draw opacity=0] (8.93,-4.29) -- (0,0) -- (8.93,4.29) -- cycle    ;

\draw   (70,205) .. controls (70,191.19) and (81.19,180) .. (95,180) .. controls (108.81,180) and (120,191.19) .. (120,205) .. controls (120,218.81) and (108.81,230) .. (95,230) .. controls (81.19,230) and (70,218.81) .. (70,205) -- cycle ;
\draw    (95,180) -- (95,152) ;
\draw [shift={(95,150)}, rotate = 450] [fill={rgb, 255:red, 0; green, 0; blue, 0 }  ][line width=0.75]  [draw opacity=0] (8.93,-4.29) -- (0,0) -- (8.93,4.29) -- cycle    ;

\draw   (70,45) .. controls (70,31.19) and (81.19,20) .. (95,20) .. controls (108.81,20) and (120,31.19) .. (120,45) .. controls (120,58.81) and (108.81,70) .. (95,70) .. controls (81.19,70) and (70,58.81) .. (70,45) -- cycle ;
\draw    (95,100) -- (95,72) ;
\draw [shift={(95,70)}, rotate = 450] [fill={rgb, 255:red, 0; green, 0; blue, 0 }  ][line width=0.75]  [draw opacity=0] (8.93,-4.29) -- (0,0) -- (8.93,4.29) -- cycle    ;

\draw   (150,125) .. controls (150,111.19) and (161.19,100) .. (175,100) .. controls (188.81,100) and (200,111.19) .. (200,125) .. controls (200,138.81) and (188.81,150) .. (175,150) .. controls (161.19,150) and (150,138.81) .. (150,125) -- cycle ;
\draw   (150,205) .. controls (150,191.19) and (161.19,180) .. (175,180) .. controls (188.81,180) and (200,191.19) .. (200,205) .. controls (200,218.81) and (188.81,230) .. (175,230) .. controls (161.19,230) and (150,218.81) .. (150,205) -- cycle ;
\draw    (175,180) -- (175,152) ;
\draw [shift={(175,150)}, rotate = 450] [fill={rgb, 255:red, 0; green, 0; blue, 0 }  ][line width=0.75]  [draw opacity=0] (8.93,-4.29) -- (0,0) -- (8.93,4.29) -- cycle    ;

\draw   (150,45) .. controls (150,31.19) and (161.19,20) .. (175,20) .. controls (188.81,20) and (200,31.19) .. (200,45) .. controls (200,58.81) and (188.81,70) .. (175,70) .. controls (161.19,70) and (150,58.81) .. (150,45) -- cycle ;
\draw    (175,100) -- (175,72) ;
\draw [shift={(175,70)}, rotate = 450] [fill={rgb, 255:red, 0; green, 0; blue, 0 }  ][line width=0.75]  [draw opacity=0] (8.93,-4.29) -- (0,0) -- (8.93,4.29) -- cycle    ;

\draw   (230,125) .. controls (230,111.19) and (241.19,100) .. (255,100) .. controls (268.81,100) and (280,111.19) .. (280,125) .. controls (280,138.81) and (268.81,150) .. (255,150) .. controls (241.19,150) and (230,138.81) .. (230,125) -- cycle ;
\draw   (230,205) .. controls (230,191.19) and (241.19,180) .. (255,180) .. controls (268.81,180) and (280,191.19) .. (280,205) .. controls (280,218.81) and (268.81,230) .. (255,230) .. controls (241.19,230) and (230,218.81) .. (230,205) -- cycle ;
\draw    (255,180) -- (255,152) ;
\draw [shift={(255,150)}, rotate = 450] [fill={rgb, 255:red, 0; green, 0; blue, 0 }  ][line width=0.75]  [draw opacity=0] (8.93,-4.29) -- (0,0) -- (8.93,4.29) -- cycle    ;

\draw   (230,45) .. controls (230,31.19) and (241.19,20) .. (255,20) .. controls (268.81,20) and (280,31.19) .. (280,45) .. controls (280,58.81) and (268.81,70) .. (255,70) .. controls (241.19,70) and (230,58.81) .. (230,45) -- cycle ;
\draw    (255,100) -- (255,72) ;
\draw [shift={(255,70)}, rotate = 450] [fill={rgb, 255:red, 0; green, 0; blue, 0 }  ][line width=0.75]  [draw opacity=0] (8.93,-4.29) -- (0,0) -- (8.93,4.29) -- cycle    ;

\draw    (200,125) -- (228,125) ;
\draw [shift={(230,125)}, rotate = 180] [fill={rgb, 255:red, 0; green, 0; blue, 0 }  ][line width=0.75]  [draw opacity=0] (8.93,-4.29) -- (0,0) -- (8.93,4.29) -- cycle    ;

\draw    (280,125) -- (338,125) ;
\draw [shift={(340,125)}, rotate = 180] [fill={rgb, 255:red, 0; green, 0; blue, 0 }  ][line width=0.75]  [draw opacity=0] (8.93,-4.29) -- (0,0) -- (8.93,4.29) -- cycle    ;

\draw   (340,125) .. controls (340,111.19) and (351.19,100) .. (365,100) .. controls (378.81,100) and (390,111.19) .. (390,125) .. controls (390,138.81) and (378.81,150) .. (365,150) .. controls (351.19,150) and (340,138.81) .. (340,125) -- cycle ;
\draw   (340,205) .. controls (340,191.19) and (351.19,180) .. (365,180) .. controls (378.81,180) and (390,191.19) .. (390,205) .. controls (390,218.81) and (378.81,230) .. (365,230) .. controls (351.19,230) and (340,218.81) .. (340,205) -- cycle ;
\draw    (365,180) -- (365,152) ;
\draw [shift={(365,150)}, rotate = 450] [fill={rgb, 255:red, 0; green, 0; blue, 0 }  ][line width=0.75]  [draw opacity=0] (8.93,-4.29) -- (0,0) -- (8.93,4.29) -- cycle    ;

\draw   (340,45) .. controls (340,31.19) and (351.19,20) .. (365,20) .. controls (378.81,20) and (390,31.19) .. (390,45) .. controls (390,58.81) and (378.81,70) .. (365,70) .. controls (351.19,70) and (340,58.81) .. (340,45) -- cycle ;
\draw    (365,100) -- (365,72) ;
\draw [shift={(365,70)}, rotate = 450] [fill={rgb, 255:red, 0; green, 0; blue, 0 }  ][line width=0.75]  [draw opacity=0] (8.93,-4.29) -- (0,0) -- (8.93,4.29) -- cycle    ;

\draw (95,205) node [scale=0.8] [align=left] {$\displaystyle \boldsymbol{x}( 0)$};
\draw (95,125) node [scale=0.8] [align=left] {$\displaystyle \boldsymbol{h}( 0)$};
\draw (95,45) node [scale=0.8] [align=left] {$\displaystyle \boldsymbol{o}( 0)$};
\draw (175,205) node [scale=0.8] [align=left] {$\displaystyle \boldsymbol{x}( 1)$};
\draw (175,125) node [scale=0.8] [align=left] {$\displaystyle \boldsymbol{h}( 1)$};
\draw (175,45) node [scale=0.8] [align=left] {$\displaystyle \boldsymbol{o}( 1)$};
\draw (255,205) node [scale=0.8] [align=left] {$\displaystyle \boldsymbol{x}( 2)$};
\draw (255,125) node [scale=0.8] [align=left] {$\displaystyle \boldsymbol{h}( 2)$};
\draw (255,45) node [scale=0.8] [align=left] {$\displaystyle \boldsymbol{o}( 2)$};
\draw (311.5,201) node  [align=left] {...};
\draw (365,205) node [scale=0.8] [align=left] {$\displaystyle \boldsymbol{x}( t)$};
\draw (365,125) node [scale=0.8] [align=left] {$\displaystyle \boldsymbol{h}( t)$};
\draw (365,45) node [scale=0.8] [align=left] {$\displaystyle \boldsymbol{o}( t)$};

\end{tikzpicture}
\caption{An unfolded RNN.}
\label{fig:UNFOLDED}
\end{figure}

Recently, deep learning techniques have attracted great attention as they yield state of the art performance in many fields, such as image recognition and natural language processing \citep{Goodfellow2016}. Deep learning techniques are also becoming popular as an alternative framework for modelling web user behaviour. Recurrent Neural Networks (RNN) are a specialized class of neural networks and a popular architecture for sequence modelling. They are networks with loops in them, allowing information to persist. In the diagram presented in figure \ref{fig:FOLDED}, a state $\bs{h}$ looks at some input $\bs{x}$ and outputs a value $\bs{o}$.
A loop allows information to be passed from one step of the network to the next. In figure \ref{fig:UNFOLDED} the unfolded diagram is presented.
It appears that an RNN can be thought of as multiple copies of the same network, each passing a message to a successor. This chain shape makes the RNN architecture the obvious choice to use for sequential data. That is why a number of neural click models have been proposed, all employing RNN.\\
\cite{Borisov2016} present a neural click model for web search, aiming to predict user clicks on a SERP. Particularly the search behaviour is interpreted as a sequence of vector states. Learning consists in finding the components of the vector state to represent concepts that are useful for modelling user behaviour. 
They represent a user session with a sequence of vectors containing features of the query submitted, the item to predict, the query-item pair and the user’s interactions with the SERP in the form of clicks and skips of the items presented. They show that their model automatically learns the concepts of user behaviour that are set manually in the PGM framework, and that it further learns concepts that cannot be designed manually. 
\\
The RNN framework has been found useful in other web modelling areas as well. \cite{ZhangYuyu2014} introduce a neural click model for sponsored search. Zang et al. consider each user's click history as a sequence, and use the properties of RNN to include features correlated to user's current behaviour as well as sequential information of user's previous behaviours, in order to predict ad clicks. They consider each user’s ad browsing history and construct the input as features of ad impressions such as ad display position, features of the user such as queries submitted and sequential features such as time spent on landing pages of clicked ads. 

\pdfbookmark[1]{E-commerce}{e-commerce}
\section{E-commerce}\label{sec:e-commerce}
A great deal of related work in e-commerce concerns Recommender Systems (RSs). RSs are algorithms and techniques providing suggestions for items interesting to a user \citep{Jannach2010}. In e-commerce the suggestions relate to the decision-making process of what products to buy. RSs for e-commerce are primarily directed towards websites offering a wide range of products, as their main goals are to increase number of products sold and help users to discover relevant products.\\
The simplest RS is a non-personalized algorithm that recommends just the most popular items. It is a simple way to improve over a random recommendation but is not typically addressed by RS research. Most RSs personalizes the recommendations by use of items that a user has interacted with in the past, such as items previously viewed, purchased and/or rating given to those items. These RSs make use of either or both collaborative filtering and content-based filtering.\\
Collaborative filtering assumes that people who agreed in the past will agree in the future, as this approach recommends items to a user that other users with similar preferences liked in the past. Content-based filtering approaches utilize properties of the items such as category or price, to recommend additional items with similar properties.\\
Each type of system has its strengths and weaknesses. Collaborative filtering requires a large amount of information about a user’s taste to make accurate recommendations. Therefore, a cold start problem, where there is not enough information to make a recommendation, is common in collaborative filtering systems. Whereas the content-based approach needs very little information to start but is far more limited in scope, as it can only make recommendations that are similar to the original seed.

\pdfbookmark[1]{Summary}{Summary}
\section{Summary}\label{sec:summary}
Most prior work concerns user behaviour on a SERP. The state of the art in this area is the advanced probabilistic models, like the TCM and the WPC model, together with the neural click models. These models are leading as they capture most aspects of user behaviour. However, they differ when it comes to generalizing and applying these models to other click modelling areas than web search. The neural click models are the ones most capable of that. The reason lies in the way these models learn dependencies between events. While the probabilistic click models are based on findings from laboratory analyses of user's web search processes, the neural click models learn the structure directly from data.\\
The task in this thesis is to model web user behaviour on an entire website where several web pages are entered sequentially, and multiple clicks, which appear on different pages, are realised during a single user task. Furthermore, the items presented on a page consist of multiple elements such as blocks, sidebars, pop-ups etc. Our model must be able to handle this.
Also the purpose is important for the choice of model. In this thesis we use a click-based model to predict a user's intention rather than describe or predict user behaviour. It makes sense to base click probabilities on examination and attractiveness parameters, but there is no reason to believe that a user's intention depends on location of items or users' examination direction on a web page. \\
Regarding the prior work in e-commerce, predicting a user’s intent to purchase is a different task than the content ranking of a RS. Users who only click and never purchase within a session and users who click and also purchase within a session can appear to have very similar preferences for the following reason. 
Even though a user has no intention to purchase, the user will often click on a product during browsing as there is no cost to do so.
Accordingly, features of products that a user has interacted with in the past are not the obvious input to use in the task of predicting purchase intent. 


\chapter{Data}\label{chap:data}
Based on collected click log data the problem is to model user behaviour on an entire website with the purpose of predicting purchase intent. Users interact with the website by navigating on its web pages. Furthermore, they interact with the various web pages through clicks like form filling or clicks on anchor links. 

\pdfbookmark[1]{Data Set}{data}
\section{Data Set}\label{sec:data}
\begin{figure}[]
\centering
\includegraphics[scale=0.65]{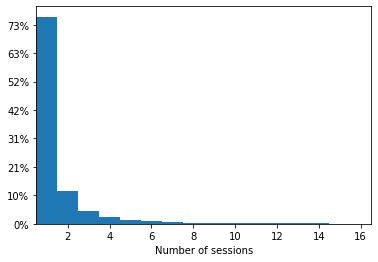}
\caption{Distribution of sessions per user.}
\label{fig:sessions}
\end{figure}

The raw click log data for this thesis consists of user IDs with URLs of visited web pages, web page timestamps and clicks within web pages. The data is logged from the website of Alka Forsikring, a Danish insurance company, in the period from May 1, 2017 to April 30, 2018, and we collected it during September 2019.\\
A user session is defined as a user visit on the website. We decide to represent a session as the presence of a user who has not visited the site anytime within the past 30 minutes.
In figure \ref{fig:sessions} a distribution of sessions per user is illustrated.
It appears that the majority of users only has one session, and that the variance is large. With this representation the sessions are more homogeneous as the session lengths vary less.
\\\\
We define the prediction target as the occurrence of a purchase in the session. Thus the prediction target indicates immediate purchase intent, meaning a user's intention to purchase in the current session rather than the user's intention to eventually purchase.\\
Only sessions where the user interacts with web pages from the e-commerce section of the website are selected, as only those sessions are exposed to purchase.\\
The collection consists of 433,141 sessions of which a purchase occurs in 13.05\%. 

\pdfbookmark[1]{Preliminary Analysis}{preanalysis}
\section{Preliminary Analysis}\label{sec:preanalysis}
\begin{figure}[]
\centering
\includegraphics[scale=0.65]{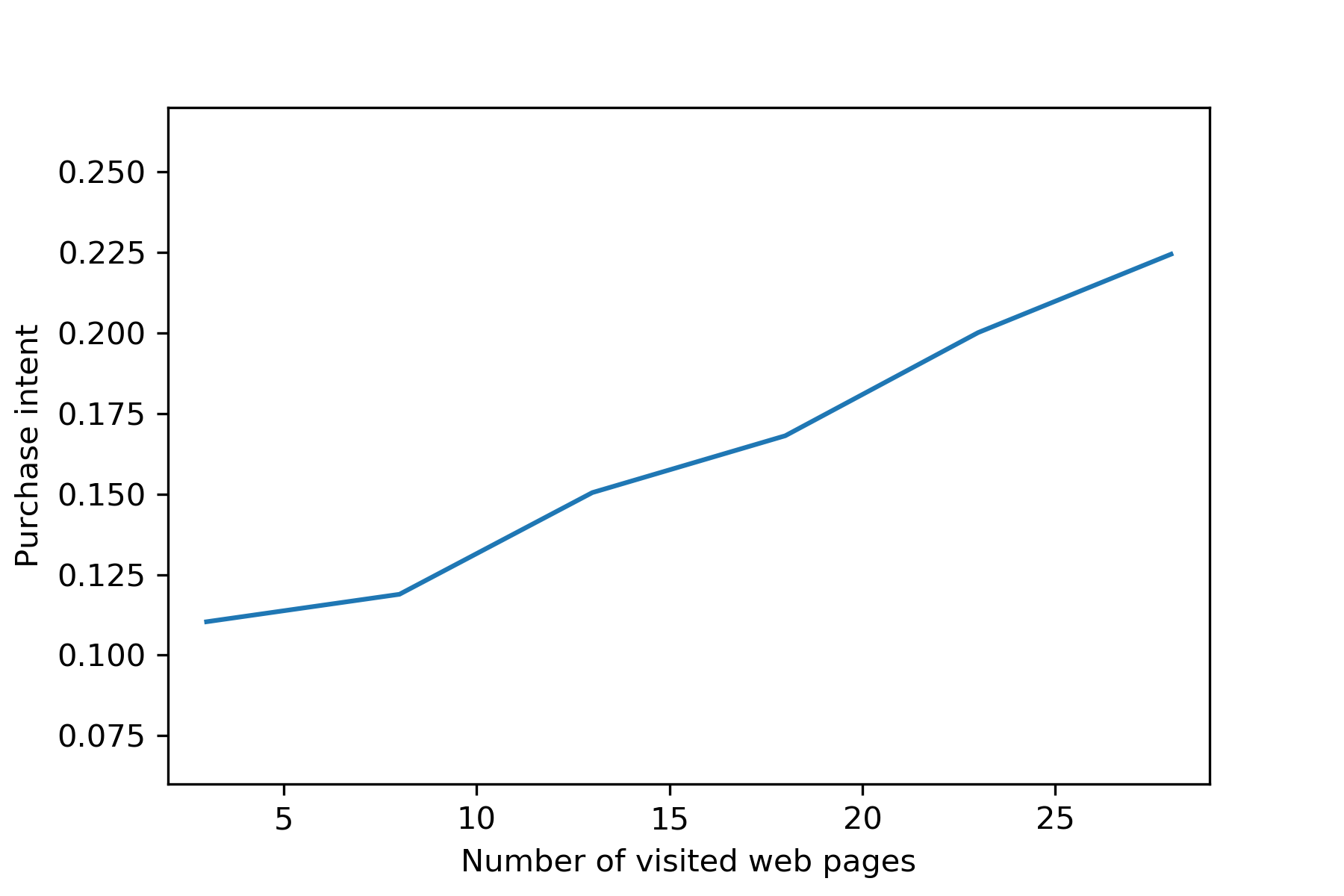}
\caption{Relationship between purchase intent and number of visited web pages.}
\label{fig:Number}
\end{figure}

\begin{figure}[]
\centering
\includegraphics[scale=0.65]{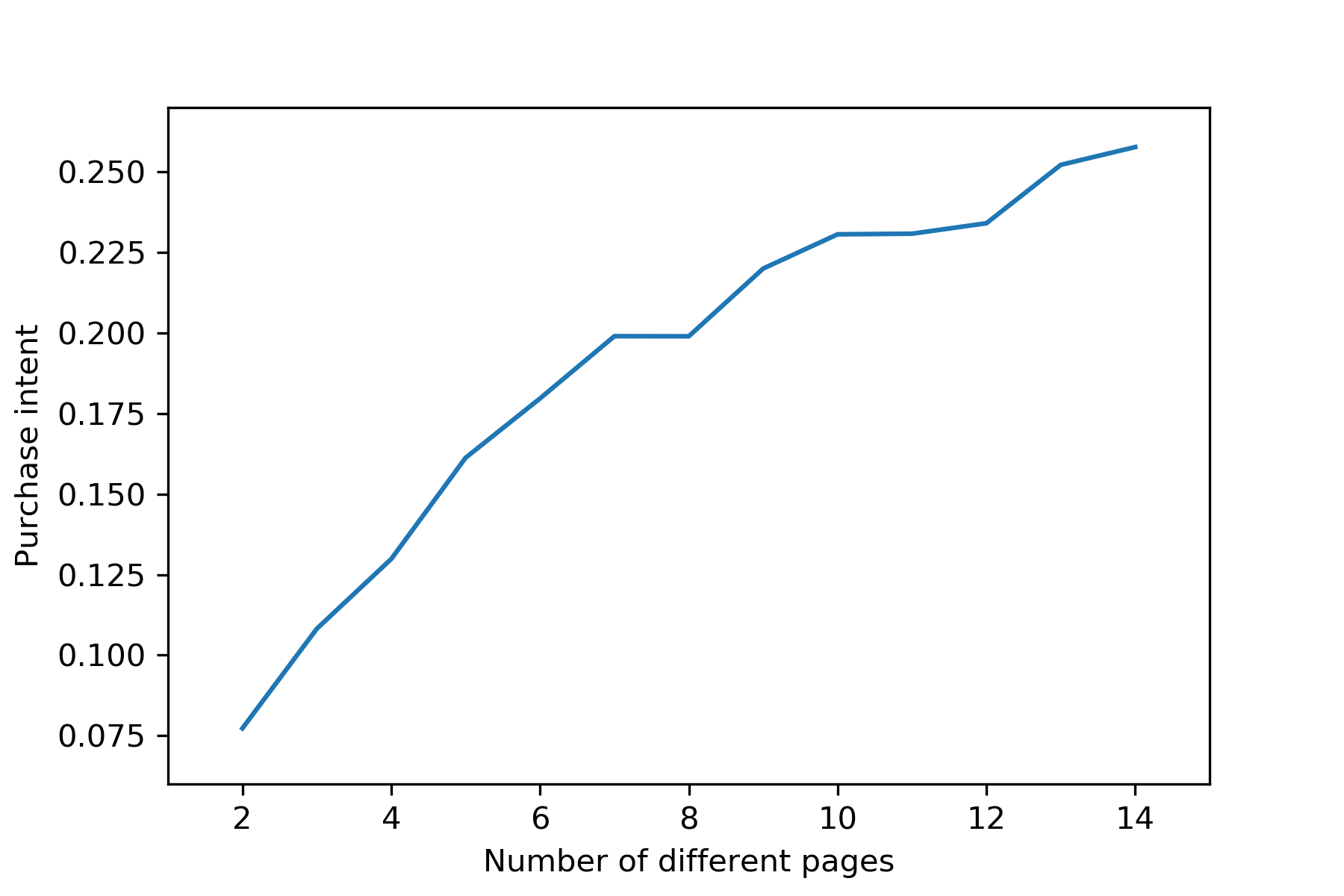}
\caption{Relationship between purchase intent and number of different web pages.}
\label{fig:Diff}
\end{figure}

\begin{figure}[]
\centering
\includegraphics[scale=0.65]{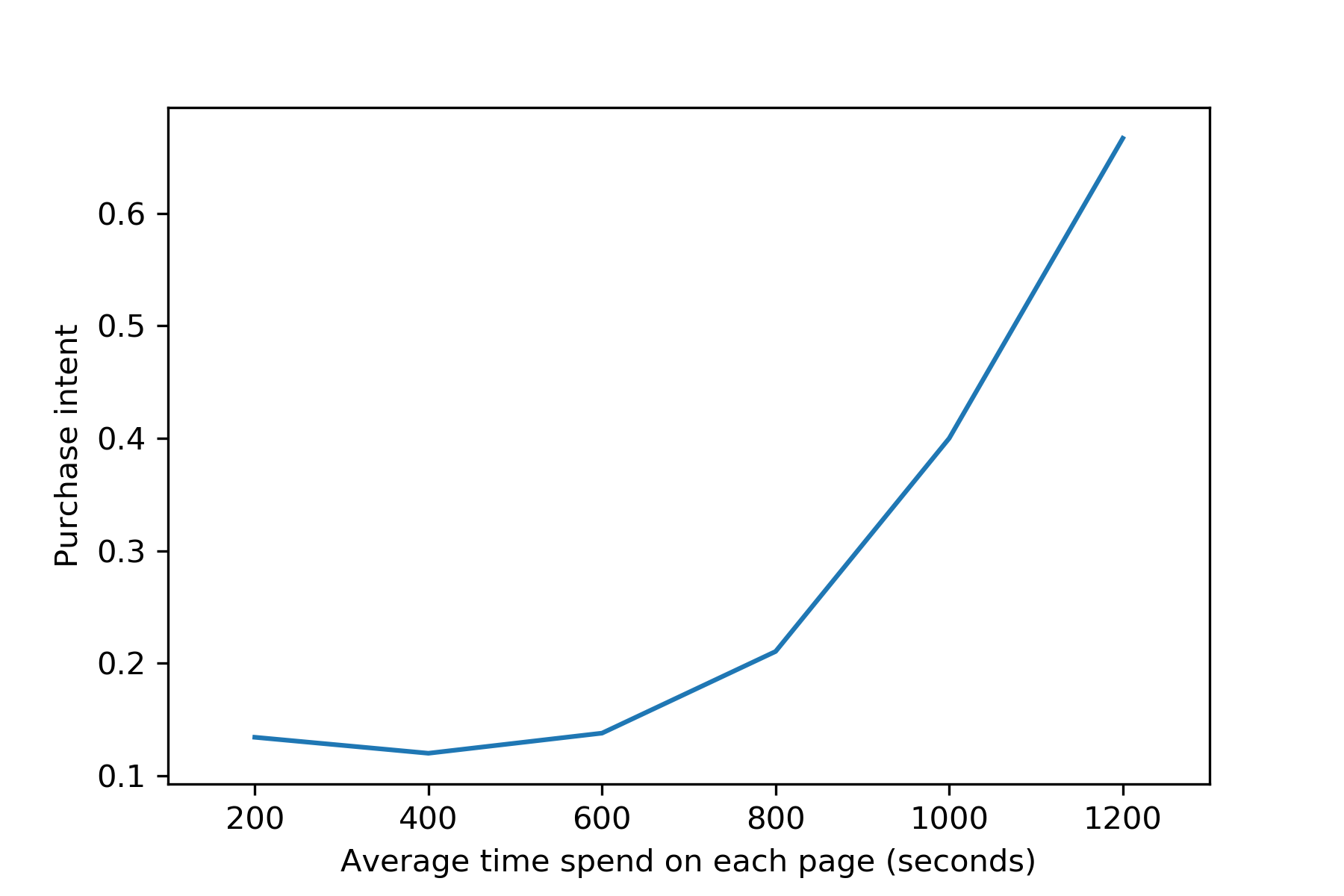}
\caption{Relationship between purchase intent and the average time spent on each web page.}
\label{fig:Time}
\end{figure}

\begin{figure}[]
\centering
\includegraphics[scale=0.65]{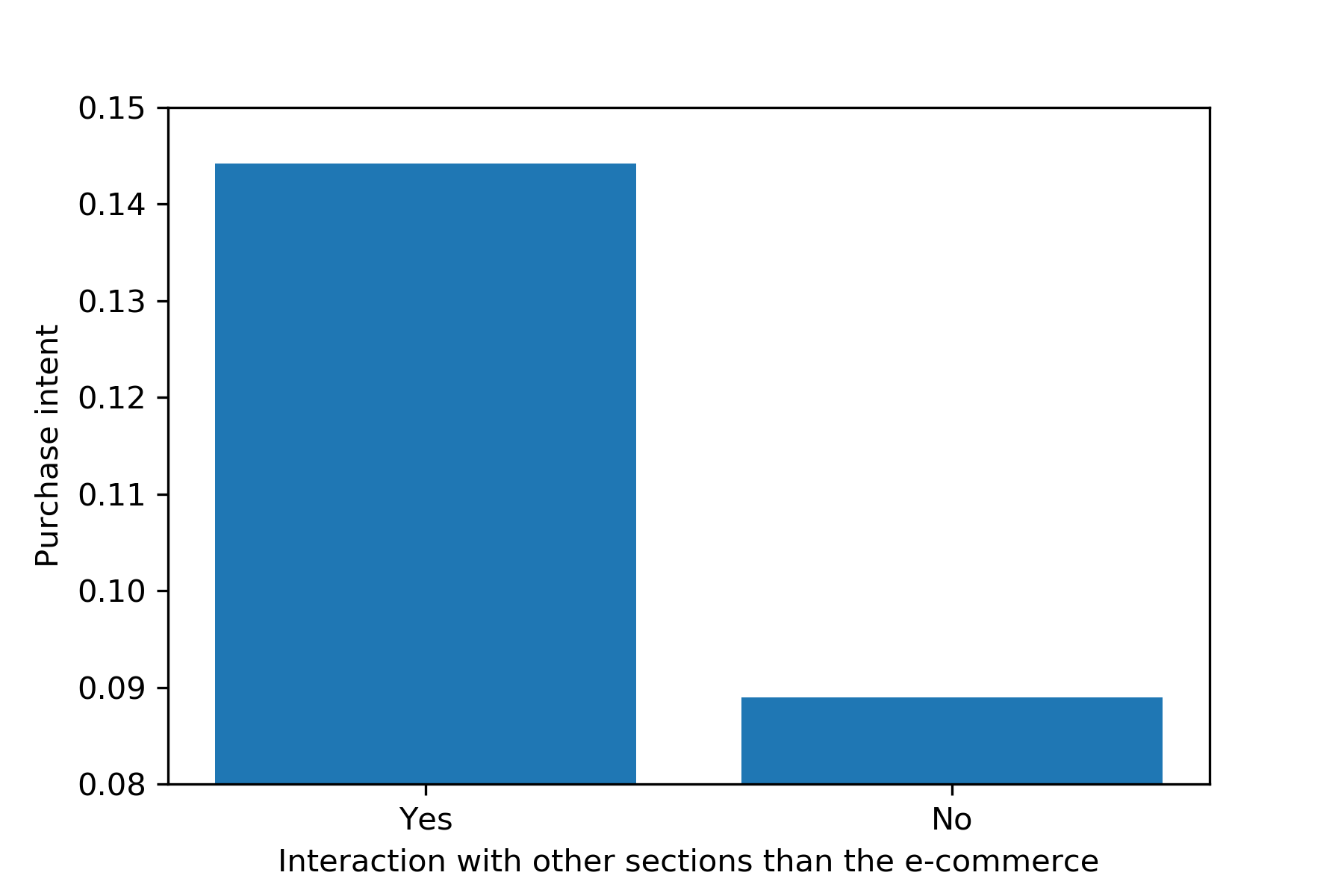}
\caption{Relationship between purchase intent and interaction with other sections than the e-commerce.}
\label{fig:Other}
\end{figure}

\begin{figure}[]
\centering
\subfloat[Car insurance.]{{\includegraphics[width=.5\textwidth]{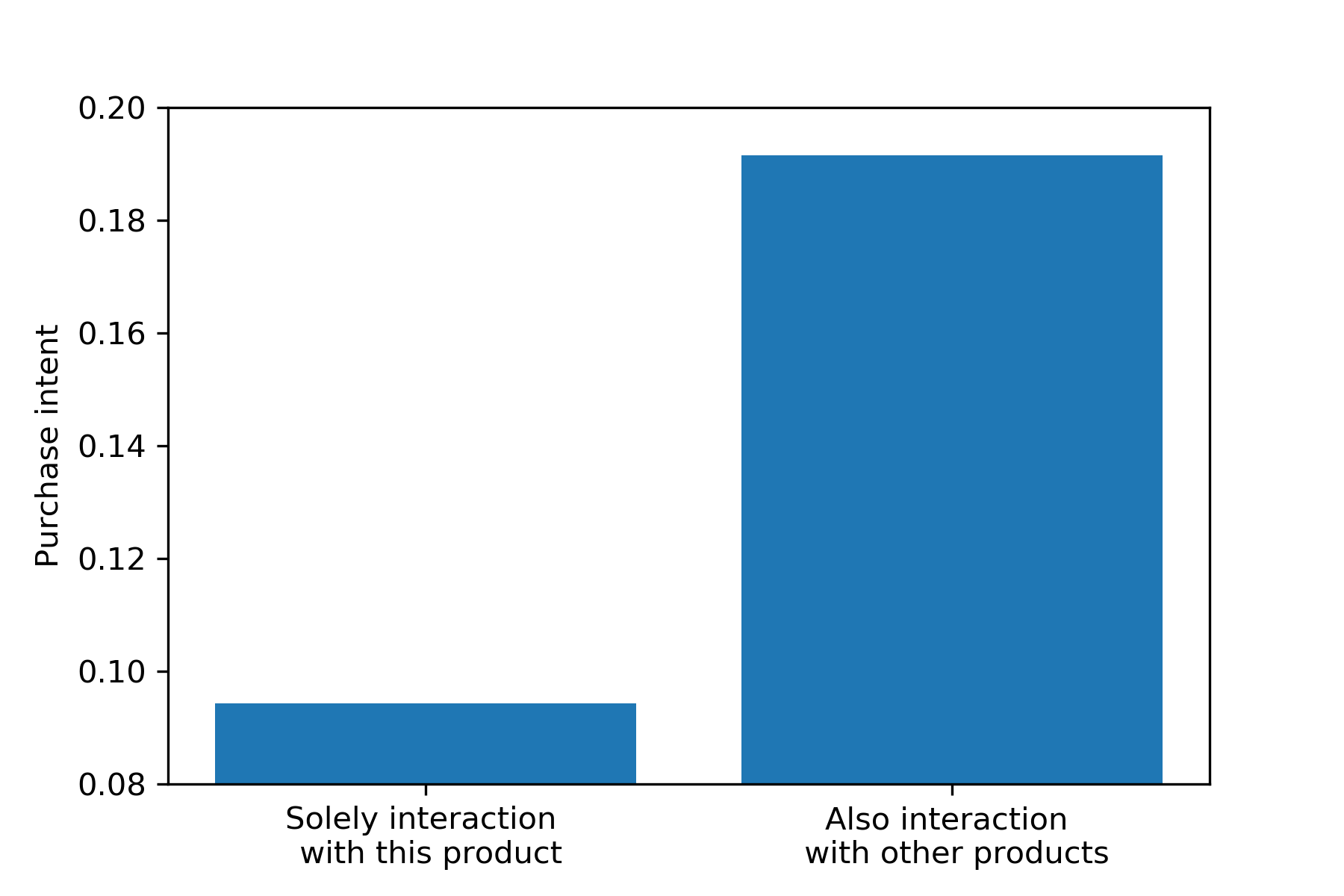}}}
\subfloat[House insurance.]{{\includegraphics[width=.5\textwidth]{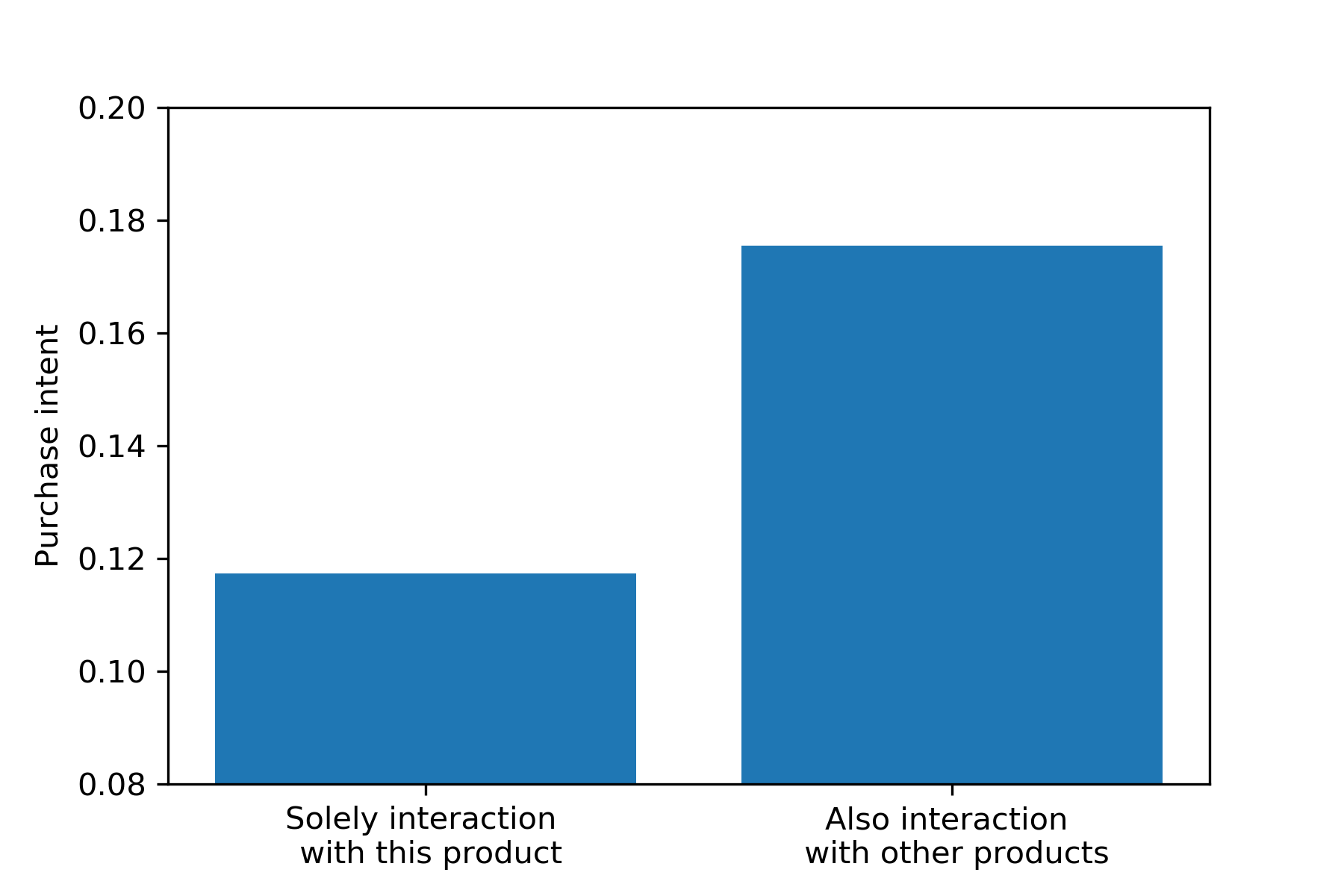}}}
\caption{Relationship between purchase intent and interactions with web pages concerning car insurance and house insurance.}
\label{fig:CarHouse}
\end{figure}

To learn how we best model user behaviour for this specific task, a preliminary analysis is conducted, where we investigate relationships between users' interactions with the website and their purchase intent within a session.\\
The main elements of a user session are the visited web pages on the website. Generally, more visited web pages implies longer sessions. To explore the effect of more visited web pages, we calculate the percentage of purchases for sessions grouped by the number of visited web pages. The correlation between number of web pages and purchase intent is shown in figure \ref{fig:Number}.
From this figure it is clear to observe a positive correlation, i.e. the more visited web pages, the more likely the user will purchase.\\
The number of visited web pages can involve many visits on the same web page. Figure \ref{fig:Diff} shows an even more significant dependency between users' purchase intent and the number of different web pages visited within the website.
\\
Our above analysis also indicates that longer sessions can give rise to rather greater purchase intent. However, it does not distinguish between users rushing through many web pages, and users spending more time on each web page during the session. We explore the relationship between purchase intent and the average time spent on each web page on the website. This is illustrated in figure \ref{fig:Time}. 
The purchase intent starts steadily in a certain level and then grows significantly, which implies that from a certain point the time aspect has an influence on purchase intent.\\
Besides our studies on the effect of number and duration of user interactions, we do a further analysis of interactions with different categories of web pages. Due to the prediction task, user interactions with the e-commerce section of the website are of greatest interest, but some of the sessions contain interactions with web pages from other sections of the website as well. To explore the effect of this, the percentage of purchases is compared for sessions with and without interactions with other sections than the e-commerce section. Figure \ref{fig:Other} shows the result. 
Furthermore, we categorize the interactions with web pages from the e-commerce section into five categories based on the products the web pages concern. The categories refer to the four main products of the insurance company, car insurance, contents insurance, house insurance and accident insurance, and then a united category for the small products like dog insurance and caravan insurance.
\begin{figure}[]
\centering
\includegraphics[scale=0.65]{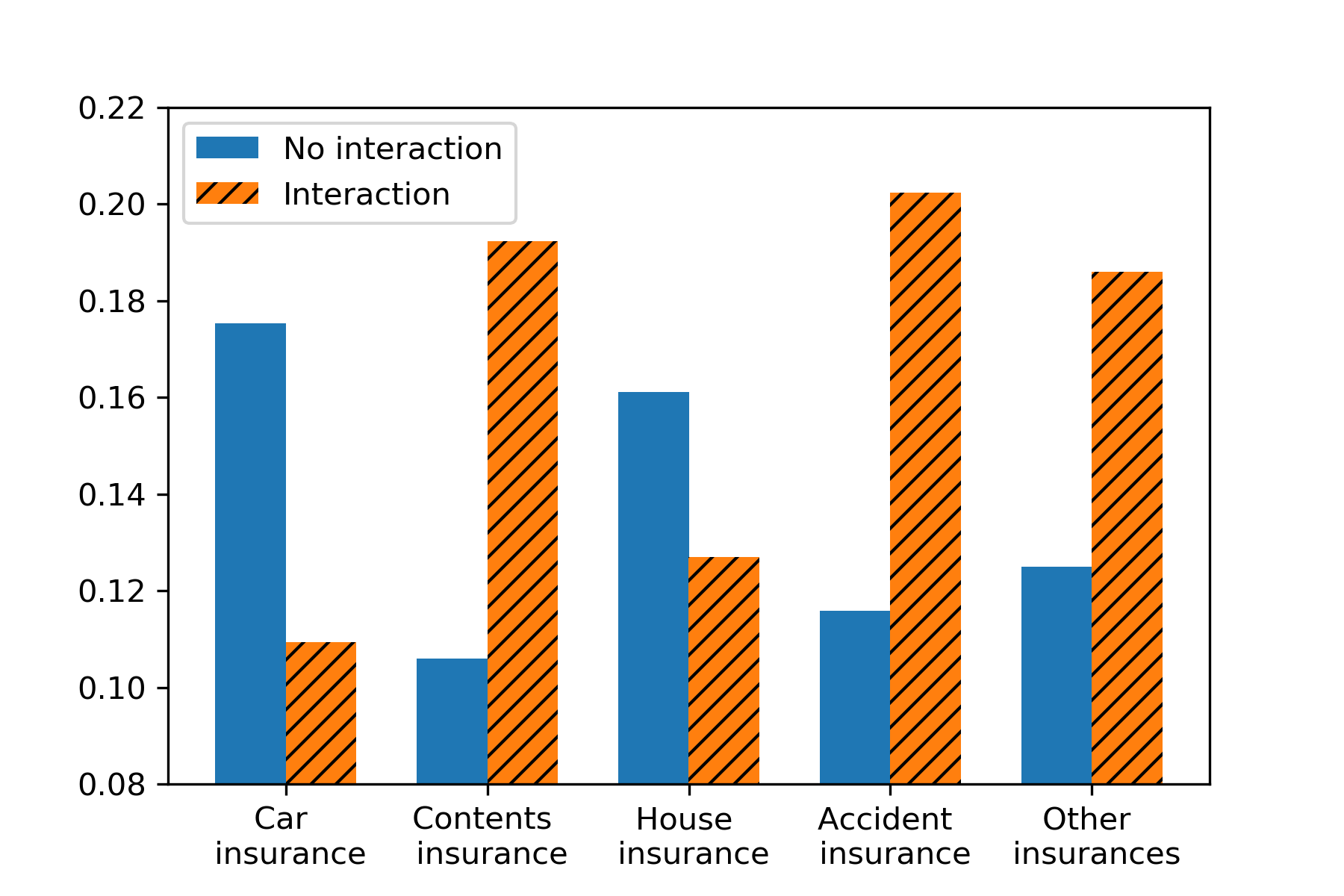}
\caption{Relationship between purchase intent and interactions with web pages concerning different insurance products.}
\label{fig:Products}
\end{figure}
From figure \ref{fig:Products} it appears that users interacting with web pages concerning car insurance and house insurance have lower purchase intent than users who do not. As regards the other insurance products, it is the other way around. However, having a deeper look into user interactions with car insurance and house insurance, we see that the purchase intent strongly depends on whether the user solely interacts with the products, or if the user also interacts with other products. This is illustrated in figure \ref{fig:CarHouse}.
%
%
\\
Figures \ref{fig:Other}, \ref{fig:Products} and \ref{fig:CarHouse} illustrate big differences in purchase intent when grouping the sessions by their interactions with different web pages. 
\\\\
All the analysis results above indicate that users' historical interactions with the website contain valuable information for the task of predicting purchase intent. Even though it is possible to identify such dependencies, it is a big challenge to enumerate them in the data manually, and it is preferable to choose a model with the ability to learn such kinds of dependencies by itself. To this end, we propose to use a feature-based framework to model user behaviour, where interactions from the entire user session including all web pages, clicks and time are represented by features. Then we can use a supervised model to learn a mapping between these features and the prediction target.
In this way the user behaviour is directly modeled via the dependency on users' historical interactions with the entire website with no need of predefined assumptions. 

\chapter{Modelling User Behaviour}\label{chap:models}
Given a user session, the aim is to predict a particular event indicating the user's intention.
We propose a feature-based framework for this task.
In the following, two different feature models are presented. 
First an \textit{engineered} click model where we engineer features of a user session into a single vector, then a \textit{sequential} click model\footnote{Note that there is no relation between this model and the Partially Sequential Click Model introduced by \cite{Wang2015}.} where we represent the features with a sequence of vectors. 


\pdfbookmark[1]{Engineered Click Model}{engineered model}
\section{Engineered Click Model}\label{sec:engineered model}
\begin{figure}[]
\centering
\scalebox{0.99}{\import{Images/}{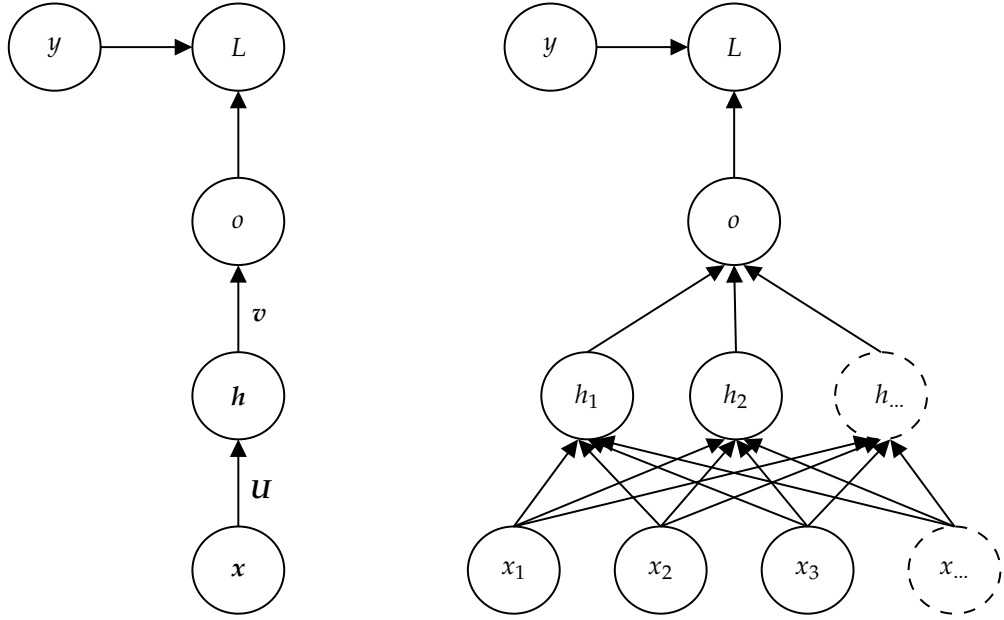}}
\caption{An example of an FFNN drawn in two different styles. Left: A compact style with one node representing each layer. The edges are annotated with the name of the parameters that describe the relationship between two layers. Right: Every unit is drawn as a node in the graph. Note that intercept parameters are omitted.}
\label{fig:FFNN}
\end{figure}

In this framework we represent features of a user session with a single vector.
Let $\bs{x}$ be the feature vector, and let $y$ be the prediction target. Since the task is to predict a single event, $y$ is a binary value indicating whether the event has occurred or not. The vector $\bs{x}$ can be the input in a variety of classification models using $y$ as prediction target. Later on, we are going to evaluate the performance of this model in relation to an RNN model. To this end, we propose a Feed Forward Neural Network (FFNN) in order to use a model with the same ability to capture interactions between features and non-linear relationships as the RNN.\\
The network must map the input vector to a single output. The mapping can be designed in different ways \citep{Goodfellow2016}, take for instance an FFNN with one hidden layer. See figure \ref{fig:FFNN} for an illustration of this model. 
The model takes the input $\bs{x}$ and computes a vector $\bs{h}$ of hidden units. The values of these hidden units are then used as the input for the output unit $o$ of the network. A loss function $L$ takes the output $o$ and the true label $y$ as inputs. The forward pass proceeds as follows.
\begin{align}
\bs{h} &= f(\bs{a}+\bs{U}\bs{x}) \\
o &= \bs{b}+\bs{h}^T\bs{v}  \\
\hat{y} &=\sigma(o) \label{equ:yhat1},
\end{align}
where the parameters are the bias vectors $\boldsymbol{a}$ and $\boldsymbol{b}$ along with the weight matrix $\boldsymbol{U}$ and weight vector $\boldsymbol{v}$.
The activation function for the hidden units could be any non-linear and differentiable function, for instance the hyperbolic tangent function, i.e. $f(z) = \frac{1-e^{-2z}}{1+e^{-2z}}$. In order to predict the probability of an event, we choose the activation function for the output layer to be the sigmoid function, $\sigma(z) = \frac{1}{1+e^{-z}}$, because its output range from 0 to 1.\\
We choose to define the loss function as the cross-entropy. This is a typical choice, when the problem is to predict a binary event, because of it's relation to the log-likelihood function of Bernoulli distributed random variables. Thus the function becomes
\begin{align}
L = -  \Big( y \log ( \hat{y} )+(1-y)\log( 1-\hat{y} ) \Big), 
\end{align}
where $\hat{y}$ are as defined in (\ref{equ:yhat1}).\\
The non-linearity of the network causes the loss function to become non-convex. Thus it should be trained with an iterative, gradient-based optimizer. We obtain the gradient with respect to the parameters using the efficient and exact Back Propagation algorithm \citep{Goodfellow2016}.
\pdfbookmark[1]{Feature Engineering}{engineering}
\subsection{Feature Engineering}\label{subsec:engineering}
Our main assumption in this framework is that features of a user's session history have predictive power, independent of the temporal order. However, since sessions can have a varying number of interactions with the website, it is not necessarily straightforward how to define such features. For instance, the time between user interactions may affect the user's intention but cannot readily be included. Thus we compute the engineered features using different aggregate functions. The aim is to transform the data into a tabular format with the least possible loss of information. Then the model is supposed to learn dependencies like the ones we saw in the preliminary analysis in section \ref{sec:preanalysis}. \\
We sum the total time a user spent on each web page during the session. Moreover, to account for sessions with varying length, we compute a ratio of visits on every web page. The ratio should indicate what proportion of the visit the web page constitutes. We compute the sum of all possible clicks for each session as well. Then we divide this sum with the number of times the corresponding click field appears in the session. We include this click rate to both incorporate what the user clicked and did not click. If a click field has never appeared, we set this click rate to zero. Finally, for each session we include the total number of visited web pages as well as the average number of clicks per web page. The features are presented in table \ref{tab:engineered}.
\begin{table}[]
\centering
\begin{tabular}{@{}ll@{}}
\toprule
Feature                                    & Description                                                                                                                               \\ \midrule
\rowcolor[HTML]{EFEFEF} 
$x_{time ~ i}$                             & Total time spent on web page $i$.                                                                                                         \\
$x_{pages}$                                & Total number of visited web pages.                                                                                                        \\
\rowcolor[HTML]{EFEFEF} 
$\frac{x_{page ~ i}}{x_{pages}}$           & \begin{tabular}[c]{@{}l@{}}Ratio of web page $i$, where $x_{page ~ i}$  is the\\ sum of visits on web page $i$.\end{tabular}              \\
$x_{click ~ j}$                            & Sum of click $j$.                                                                                                                         \\
\rowcolor[HTML]{EFEFEF} 
$\frac{x_{click ~ j}}{x_{appearance ~ j}}$ & \begin{tabular}[c]{@{}l@{}}Rate of click $j$, where $x_{appearance ~ j}$ is the \\ number of appearances of click field $j$.\end{tabular} \\
$x_{avg ~ clicks}$                         & Average number of clicks per web page.                                                                                                    \\ \bottomrule
\end{tabular}
\caption{For each session a list of features is computed.}
\label{tab:engineered}
\end{table}

\pdfbookmark[1]{Sequential Click Model}{sequential model}
\section{Sequential Click Model}\label{sec:sequential model}
\begin{figure}[]
\centering
\tikzset{every picture/.style={line width=0.75pt}} 

\begin{tikzpicture}[x=0.75pt,y=0.75pt,yscale=-1,xscale=1]

\draw  [dash pattern={on 4.5pt off 4.5pt}] (30,265) .. controls (30,251.19) and (41.19,240) .. (55,240) .. controls (68.81,240) and (80,251.19) .. (80,265) .. controls (80,278.81) and (68.81,290) .. (55,290) .. controls (41.19,290) and (30,278.81) .. (30,265) -- cycle ;
\draw   (110,265) .. controls (110,251.19) and (121.19,240) .. (135,240) .. controls (148.81,240) and (160,251.19) .. (160,265) .. controls (160,278.81) and (148.81,290) .. (135,290) .. controls (121.19,290) and (110,278.81) .. (110,265) -- cycle ;
\draw    (80,265) -- (108,265) ;
\draw [shift={(110,265)}, rotate = 180] [fill={rgb, 255:red, 0; green, 0; blue, 0 }  ][line width=0.75]  [draw opacity=0] (8.93,-4.29) -- (0,0) -- (8.93,4.29) -- cycle    ;

\draw   (110,345) .. controls (110,331.19) and (121.19,320) .. (135,320) .. controls (148.81,320) and (160,331.19) .. (160,345) .. controls (160,358.81) and (148.81,370) .. (135,370) .. controls (121.19,370) and (110,358.81) .. (110,345) -- cycle ;
\draw    (135,320) -- (135,292) ;
\draw [shift={(135,290)}, rotate = 450] [fill={rgb, 255:red, 0; green, 0; blue, 0 }  ][line width=0.75]  [draw opacity=0] (8.93,-4.29) -- (0,0) -- (8.93,4.29) -- cycle    ;

\draw   (190,265) .. controls (190,251.19) and (201.19,240) .. (215,240) .. controls (228.81,240) and (240,251.19) .. (240,265) .. controls (240,278.81) and (228.81,290) .. (215,290) .. controls (201.19,290) and (190,278.81) .. (190,265) -- cycle ;
\draw   (190,345) .. controls (190,331.19) and (201.19,320) .. (215,320) .. controls (228.81,320) and (240,331.19) .. (240,345) .. controls (240,358.81) and (228.81,370) .. (215,370) .. controls (201.19,370) and (190,358.81) .. (190,345) -- cycle ;
\draw    (160,265) -- (188,265) ;
\draw [shift={(190,265)}, rotate = 180] [fill={rgb, 255:red, 0; green, 0; blue, 0 }  ][line width=0.75]  [draw opacity=0] (8.93,-4.29) -- (0,0) -- (8.93,4.29) -- cycle    ;

\draw    (215,320) -- (215,292) ;
\draw [shift={(215,290)}, rotate = 450] [fill={rgb, 255:red, 0; green, 0; blue, 0 }  ][line width=0.75]  [draw opacity=0] (8.93,-4.29) -- (0,0) -- (8.93,4.29) -- cycle    ;

\draw  [dash pattern={on 4.5pt off 4.5pt}] (270,265) .. controls (270,251.19) and (281.19,240) .. (295,240) .. controls (308.81,240) and (320,251.19) .. (320,265) .. controls (320,278.81) and (308.81,290) .. (295,290) .. controls (281.19,290) and (270,278.81) .. (270,265) -- cycle ;
\draw  [dash pattern={on 4.5pt off 4.5pt}] (270,345) .. controls (270,331.19) and (281.19,320) .. (295,320) .. controls (308.81,320) and (320,331.19) .. (320,345) .. controls (320,358.81) and (308.81,370) .. (295,370) .. controls (281.19,370) and (270,358.81) .. (270,345) -- cycle ;
\draw    (240,265) -- (268,265) ;
\draw [shift={(270,265)}, rotate = 180] [fill={rgb, 255:red, 0; green, 0; blue, 0 }  ][line width=0.75]  [draw opacity=0] (8.93,-4.29) -- (0,0) -- (8.93,4.29) -- cycle    ;

\draw    (295,320) -- (295,292) ;
\draw [shift={(295,290)}, rotate = 450] [fill={rgb, 255:red, 0; green, 0; blue, 0 }  ][line width=0.75]  [draw opacity=0] (8.93,-4.29) -- (0,0) -- (8.93,4.29) -- cycle    ;

\draw   (350,265) .. controls (350,251.19) and (361.19,240) .. (375,240) .. controls (388.81,240) and (400,251.19) .. (400,265) .. controls (400,278.81) and (388.81,290) .. (375,290) .. controls (361.19,290) and (350,278.81) .. (350,265) -- cycle ;
\draw    (320,265) -- (348,265) ;
\draw [shift={(350,265)}, rotate = 180] [fill={rgb, 255:red, 0; green, 0; blue, 0 }  ][line width=0.75]  [draw opacity=0] (8.93,-4.29) -- (0,0) -- (8.93,4.29) -- cycle    ;

\draw   (350,345) .. controls (350,331.19) and (361.19,320) .. (375,320) .. controls (388.81,320) and (400,331.19) .. (400,345) .. controls (400,358.81) and (388.81,370) .. (375,370) .. controls (361.19,370) and (350,358.81) .. (350,345) -- cycle ;
\draw    (375,320) -- (375,292) ;
\draw [shift={(375,290)}, rotate = 450] [fill={rgb, 255:red, 0; green, 0; blue, 0 }  ][line width=0.75]  [draw opacity=0] (8.93,-4.29) -- (0,0) -- (8.93,4.29) -- cycle    ;

\draw   (350,185) .. controls (350,171.19) and (361.19,160) .. (375,160) .. controls (388.81,160) and (400,171.19) .. (400,185) .. controls (400,198.81) and (388.81,210) .. (375,210) .. controls (361.19,210) and (350,198.81) .. (350,185) -- cycle ;
\draw    (375,240) -- (375,212) ;
\draw [shift={(375,210)}, rotate = 450] [fill={rgb, 255:red, 0; green, 0; blue, 0 }  ][line width=0.75]  [draw opacity=0] (8.93,-4.29) -- (0,0) -- (8.93,4.29) -- cycle    ;

\draw   (350,105) .. controls (350,91.19) and (361.19,80) .. (375,80) .. controls (388.81,80) and (400,91.19) .. (400,105) .. controls (400,118.81) and (388.81,130) .. (375,130) .. controls (361.19,130) and (350,118.81) .. (350,105) -- cycle ;
\draw    (375,160) -- (375,132) ;
\draw [shift={(375,130)}, rotate = 450] [fill={rgb, 255:red, 0; green, 0; blue, 0 }  ][line width=0.75]  [draw opacity=0] (8.93,-4.29) -- (0,0) -- (8.93,4.29) -- cycle    ;

\draw   (270,105) .. controls (270,91.19) and (281.19,80) .. (295,80) .. controls (308.81,80) and (320,91.19) .. (320,105) .. controls (320,118.81) and (308.81,130) .. (295,130) .. controls (281.19,130) and (270,118.81) .. (270,105) -- cycle ;
\draw    (320,105) -- (348,105) ;
\draw [shift={(350,105)}, rotate = 180] [fill={rgb, 255:red, 0; green, 0; blue, 0 }  ][line width=0.75]  [draw opacity=0] (8.93,-4.29) -- (0,0) -- (8.93,4.29) -- cycle    ;

\draw (55,265) node [scale=0.8] [align=left] {$\displaystyle ...$};
\draw (135,265) node [scale=0.8] [align=left] {$\displaystyle \boldsymbol{h}( t-1)$};
\draw (91.5,251.5) node [scale=0.8] [align=left] {$\displaystyle \boldsymbol{W}$};
\draw (135,345) node [scale=0.8] [align=left] {$\displaystyle \boldsymbol{x}( t-1)$};
\draw (147.5,308.5) node [scale=0.8] [align=left] {$\displaystyle \boldsymbol{U}$};
\draw (215,345) node [scale=0.8] [align=left] {$\displaystyle \boldsymbol{x}( t)$};
\draw (215,265) node [scale=0.8] [align=left] {$\displaystyle \boldsymbol{h}( t)$};
\draw (171.5,251.5) node [scale=0.8] [align=left] {$\displaystyle \boldsymbol{W}$};
\draw (295,265) node [scale=0.8] [align=left] {$\displaystyle \boldsymbol{h}( ...)$};
\draw (295,345) node [scale=0.8] [align=left] {$\displaystyle \boldsymbol{x}( ...)$};
\draw (251.5,251.5) node [scale=0.8] [align=left] {$\displaystyle \boldsymbol{W}$};
\draw (307.5,308.5) node [scale=0.8] [align=left] {$\displaystyle \boldsymbol{U}$};
\draw (227.5,308.5) node [scale=0.8] [align=left] {$\displaystyle \boldsymbol{U}$};
\draw (331.5,251.5) node [scale=0.8] [align=left] {$\displaystyle \boldsymbol{W}$};
\draw (375,345) node [scale=0.8] [align=left] {$\displaystyle \boldsymbol{x}( \tau )$};
\draw (387.5,308.5) node [scale=0.8] [align=left] {$\displaystyle \boldsymbol{U}$};
\draw (375,265) node [scale=0.8] [align=left] {$\displaystyle \boldsymbol{h}( \tau )$};
\draw (387,228.5) node [scale=0.8] [align=left] {$\displaystyle \boldsymbol{v}$};
\draw (375,185) node [scale=0.8] [align=left] {$\displaystyle o( \tau )$};
\draw (375,105) node [scale=0.8] [align=left] {$\displaystyle L( \tau )$};
\draw (295,105) node [scale=0.8] [align=left] {$\displaystyle y( \tau )$};

\end{tikzpicture}
\caption{We propose to use an RNN that takes a sequence as input and returns a single output at the end. Thus the network has connection between $\bs{x}$ and $\bs{h}$ for every time step $t$ and only a connection between $\bs{h}$ and $o$ at the final time step $\tau$.}
\label{fig:RNN}
\end{figure}
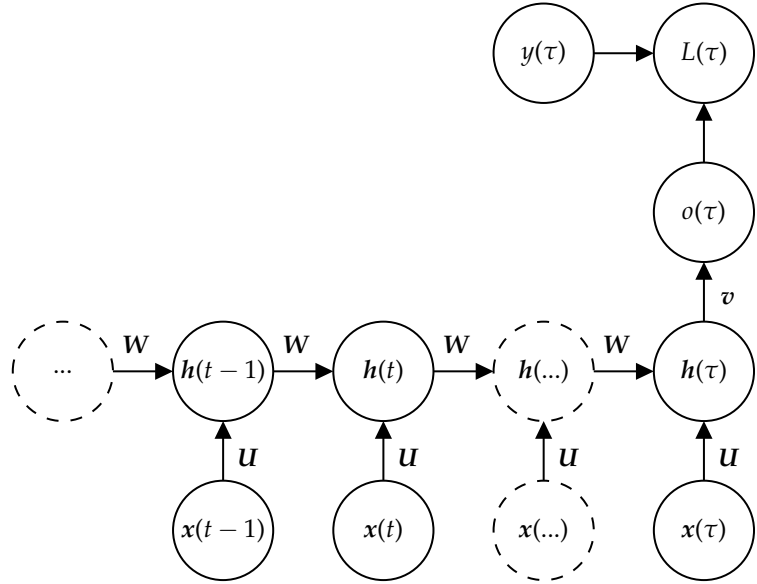

In this framework we represent features of a user session at different time steps with a sequence of vectors. 
Let $t$ denote the time step index that refers to the position in the sequence, and let $\bs{x}(t)$ be the vector at that time. Let $\tau$ be the time to be predicted, and let $y(\tau)$ be the true label. As in the framework based on feature engineering, $y(\tau)$ is a binary value. We propose to use $\bs{x}(t)$ for $t \leq \tau$ as the input vectors in an RNN with $y(\tau)$ as the the prediction target.\\
An RNN can have different design patterns \citep{Goodfellow2016}. For this kind of problem, the network should read an entire sequence with recurrent connections between hidden units and then produce a single output. The architecture of such a network is illustrated in figure \ref{fig:RNN}.
As the FFNN, it consists of an input layer $\bs{x}$, a hidden layer $\bs{h}$ and an output unit $o$, but the hidden layer now involves a recurrence relation. The forward propagation equations are now as follows. 
\begin{align}
\boldsymbol{h}(t) &= f(\boldsymbol{a}+\boldsymbol{W}\boldsymbol{h}(t-1)+\boldsymbol{U}\boldsymbol{x}(t)),~~\text{for}~~ t \leq \tau \label{equ:h1} \\
o(\tau) &= \boldsymbol{b}+\boldsymbol{h}(\tau)^T\boldsymbol{v} \\
\hat{y}(\tau) &=\sigma(o(\tau)), \label{equ:yhat2}
\end{align}
where the parameters from the FFNN are extended with the weight matrix $\boldsymbol{W}$ in the input-to-hidden connection. The activation function for the hidden units could be exactly as in the feed forward case. Again we choose the activation function for the output layer to be the sigmoid function, as the prediction task is unchanged.\\
While $\boldsymbol{x}(t)$ represents the features of current user behaviour, $\boldsymbol{h}(t)$ represents sequential information of previous user behaviour. Thus the prediction $\hat{y}(\tau)$ depends on not only the current input features but also the sequential historical information. \\
Once again we define the loss function as the cross-entropy, thus it becomes
\begin{align}
L(\tau) = -  \Big( y(\tau) \log \Big( \hat{y}(\tau) \Big)+(1-y(\tau))\log \Big( 1-\hat{y}(\tau) \Big) \Big), \\
\end{align}
where $\hat{y}$ are as defined in (\ref{equ:yhat2}).
\pdfbookmark[1]{Back Propagation Through Time}{BPTT}
\subsection{Back Propagation Through Time}\label{subsec:BPTT}
The loss function can be minimized with any gradient-based technique. However, because of the recurrent connections, we cannot compute the gradient with respect to the parameters with the standard Back Propagation algorithm as in the feed forward case \citep{Goodfellow2016}.
Instead, we apply the generalized back propagation algorithm on the nodes in the time-unfolded RNN. This is called Back Propagation Through Time (BPTT), and in this particular model the computations proceed as follows. \\
For each node we compute the gradient recursively, based on the gradient from nodes that follow it. The recursion is started with the loss. We compute the gradient of the output layer as
\begin{align}
\nabla_{o(\tau)}L &= \frac{\partial L}{\partial o(\tau)} \\
                               &= \hat{y}(\tau)-y(\tau).
\end{align}
Next the computations go backward in the sequence. At the final time step, $\tau$, $\boldsymbol{h}(\tau)$ only has $o(\tau)$ as a descendent, so its gradient is simply
\begin{align}
\nabla_{\boldsymbol{h}(\tau)}L = \boldsymbol{v}^T \nabla_{o(\tau)}L.
\end{align}
At every time step $t<\tau$, $\boldsymbol{h}(t)$ has $\boldsymbol{h}(t+1)$ as descendent. With $f$ in equation (\ref{equ:h1}) being the hyperbolic tangent function, the gradient is given by
\begin{align}
\nabla_{\boldsymbol{h}(t)}L &= \left(\frac{\partial \boldsymbol{h}(t+1)}{\partial \boldsymbol{h}(t)} \right)^T (\nabla_{\boldsymbol{h}(t+1)}L) \\
                            &=\boldsymbol{W}^T \text{diag}(1-\boldsymbol{h}(t+1)^2)(\nabla_{\boldsymbol{h}(t+1)}L),
\end{align}
where ${\text{diag}(1-\boldsymbol{h}(t+1)^2)}$ indicates the diagonal matrix with the elements of ${1-\boldsymbol{h}(t+1)^2}$ on the diagonal, and we used that the derivative of $f(z) = \tanh(z)$ is ${f'(z) = 1-f(z)^2}$.\\
Now the gradients of the parameters are ready to be computed. Because the parameters are shared across many time steps dummy variables need to be introduced. For a parameter $\boldsymbol{P}$, we define $\boldsymbol{P}(t)$ as a copy, but it is only used at time step $t$. The gradients of the parameters are now given by
\begin{align}
\nabla_{\boldsymbol{b}} L &= \left( \frac{\partial o(\tau)}{\partial \boldsymbol{b}} \right)^T \nabla_{o(\tau)}L \\
                           &= \nabla_{o(\tau)}L, \\
\nabla_{\boldsymbol{a}}L &= \sum_t \left( \frac{\partial\boldsymbol{h}(t)}{\partial\boldsymbol{a}(t)} \right)^T \nabla_{\boldsymbol{h}(t)}L \\
                           &= \sum_t \text{diag}(1-\boldsymbol{h}(t)^2)\nabla_{\boldsymbol{h}(t)}L, \\
\nabla_{\boldsymbol{v}}L   &= \left( \frac{\partial o(\tau)}{\partial \boldsymbol{v}} \right)^T \nabla_{o(\tau)}L \\
                           &= \boldsymbol{h}(\tau)^T \nabla_{o(\tau)}L, \\
\nabla_{\boldsymbol{W}}L   &= \sum_t \sum_i \left( \frac{\partial L}{\partial h_i(t)} \right) \nabla_{\boldsymbol{W}(t)}h_i(t) \\
                           &= \sum_t \text{diag}(1-\boldsymbol{h}(t)^2)(\nabla_{\boldsymbol{h}(t)}L)\boldsymbol{h}(t-1)^T,\\
\nabla_{\boldsymbol{U}}L   &= \sum_t \sum_i \left( \frac{\partial L}{\partial h_i(t)} \right) \nabla_{\boldsymbol{U}(t)}h_i(t) \\
                           &= \sum_t \text{diag}(1-\boldsymbol{h}(t)^2)(\nabla_{\boldsymbol{h}(t)}L) \boldsymbol{x}(t)^T.
\end{align}
\pdfbookmark[1]{Long Short Term Memory}{LSTM}
\subsection{Long Short Term Memory}\label{subsec:LSTM}
\begin{figure}[h]
\centering
\import{Images/}{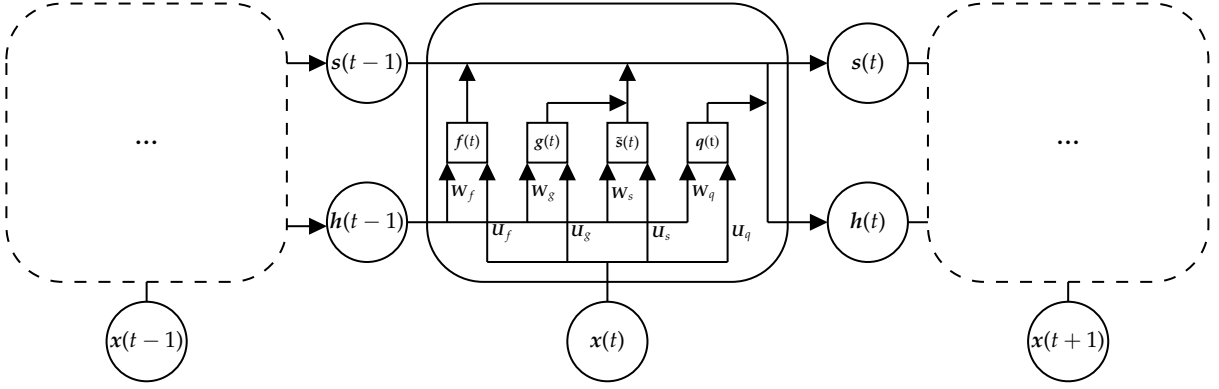}
\caption{We decide to expand the network with LSTM cells that contain four interacting layers, contrary to one layer in the recurrent cells of the standard RNN.}
\label{fig:LSTM}
\end{figure}

One of the appeals of RNNs is that they are able to connect previous information to the present task. However, there is a mathematical challenge of learning long-term dependencies in RNNs. The basic problem is that gradients propagated over many stages tend to either vanish or explode \citep{Goodfellow2016}, because RNNs repeatedly apply the same operation at each time step of a long temporal sequence. For example consider a very simple RNN with the recurrence relation
\begin{align}
\bs{h}(t) = \bs{W}\bs{h}(t-1),
\end{align}
lacking inputs and lacking a nonlinear activation function. Unfolding the equation, it simplifies to
\begin{align}
\bs{h}(t)=\bs{W}^t\bs{h}(0).
\end{align}
Suppose that $\bs{W}$ has an eigendecomposition $\bs{W}=\bs{Q}\bs{\Lambda}\bs{Q}^T$, where $\bs{Q}$ is a matrix whose $i$'th column is the eigenvector of $\bs{W}$, and $\bs{\Lambda}$ is the diagonal matrix whose diagonal elements are the corresponding eigenvalues. The recurrence relation now simplifies to
\begin{align}
\bs{h}(t)=\bs{Q}\bs{\Lambda}^t\bs{Q}^T \bs{h}(0).
\end{align}
The eigenvalues are raised to the power of $t$, causing any eigenvalues to either explode, if they are greater than 1 in magnitude, or vanish, if they are less than 1 in magnitude. All components of $\bs{h}(0)$ that are orthogonal to the principal eigenvector of $\bs{W}$ will eventually discard.\\ Long Short Term Memory (LSTM) networks are a special kind of RNN, capable of learning long-term dependencies. They are based on the idea of creating paths through time that have derivatives that neither vanish nor explode. They do it by introducing connection weights that may change at each time step. LSTM allows the network to accumulate information and to forget the old state once that information has been used. Furthermore, the neural network learns to decide when to clear the state.\\
Our RNN model does not consider the different types of relationship between users' interactions, e.g. two user interactions within a short time tend to be related, and user interactions with a large time interval may aim at different goals. LSTM is equipped with gates that are specifically designed, so that compared to the traditional RNN model, LSTM would better captures both of users short-term and long-term interests, so as to improve the model performance. Because of that, we decide to expand our proposed RNN with LSTM cells that represent the ordinary recurrent units but also contain an internal recurrence called the state unit $\bs{s}(t)$. The new LSTM cells are illustrated in figure \ref{fig:LSTM}. 
\\
We will now formalize the corresponding forward propagation equations. The first step in the LSTM is to decide what information to throw away from the cell. This decision is typically made by a sigmoid function that takes $\bs{h}(t-1)$ and $\bs{x}(t)$ as inputs and outputs a number between 0 and 1:
\begin{align}
\bs{f}(t) = \sigma(\boldsymbol{a}_f+\boldsymbol{W}_f\boldsymbol{h}(t-1)+\boldsymbol{U}_f\boldsymbol{x}(t)).
\end{align}
A value of 0 means "completely get rid of this" while a value of 1 means "completely keep this". \\
The next step is to decide what new information to store in the cell. This has two parts. A part deciding which values to update, and a part creating new candidate values, $\tilde{\bs{s}}(t)$, that could be added to the cell:
\begin{align}
\bs{g}(t) &= \sigma(\boldsymbol{a}_g+\boldsymbol{W}_g\boldsymbol{h}(t-1)+\boldsymbol{U}_g\boldsymbol{x}(t)) \\
\tilde{\bs{s}}(t) &= \tanh(\boldsymbol{a}_s+\boldsymbol{W}_s\boldsymbol{h}(t-1)+\boldsymbol{U}_s\boldsymbol{x}(t)).
\end{align}
The old unit state $\bs{s}(t-1)$ is now updated into the new unit state $\bs{s}$ by multiplying with $\bs{f}(t)$ to forget and adding the new candidate values, scaled by how much it should be updated, i.e.
\begin{align}
\bs{s}(t) = \bs{f}(t) \bs{s}(t-1)+\bs{g}(t) \tilde{\bs{s}}(t).
\end{align}
Finally, the output $\bs{h}(t)$ is computed. It consists of the unit state transformed with an activation function like the hyperbolic tangent function, and a part deciding what part of the cell to output:
\begin{align}
\bs{q}(t) &= \sigma(\boldsymbol{a}_q+\boldsymbol{W}_q\boldsymbol{h}(t-1)+\boldsymbol{U}_q\boldsymbol{x}(t)) \\
\bs{h}(t) &= \bs{q}(t) \tanh(\bs{s}(t)) \label{equ:h2}.
\end{align}
We achieve our final proposed model by replacing equation (\ref{equ:h1}) with (\ref{equ:h2}).


\chapter{Experiment}\label{chap:experiment}While click data contains indications of user preferences, most click data is also noisy. This limits the performance of the RS and wrong inference may be derived. A future direction is to examine the quality of clicks for inferring user preferences and studying methods for reducing noise in the click data.

In the research areas of web search ranking and email detection leveraging user interactions (e.g., clicks) as additional signals for learning have been extensively studied and the problem of distinguishing noisy instances from informative instances has been explored. \citet{Joachims2005} propose to interpret clicks as pairwise preference statements rather than absolute feedback. \citet{Shani2002} propose a learning mechanism based on curriculum learning and self-paced learning to judiciously select informative weak instances to learn from. We plan to study these methods for better inferring user preferences from clicks. 
\\\\
In an insurance domain it is necessary to understand whether an RS makes business sense, is compliant with laws and regulations, and can be explained to users. Since neural networks are not always inherently transparent, work on model interpretation and explanations of why items are recommended to a user is a candidate for future research. 

In recent years, there has been much research into the topic of interpretable machine learning and explainable recommendations \cite{Pepa2020, Molnar2019, Zhang2020}. We plan to study the interpretable machine learning techniques, Individual Conditional Expectation, Local Surrogate and Shapley Additive Explanations, which have flexibility to work with any machine learning model. We will focus on how to interpret the whole mapping from data input to prediction in our two-level proposed framework.
We perform two experiments. First, two models based on click data are compared with respect to the task of predicting a user's purchase intent. Secondly, the click models' ability to predict purchase intent is examined further by comparing them with a model based on demographic data.

\pdfbookmark[1]{Methods}{methods}
\section{Methods}\label{sec:methods}
The main models of the experiment are the Engineered Click Model and the Sequential Click Model, both presented in chapter \ref{chap:models}.
We implement both models with two hidden layers. The Engineered Click Model consists of a fully connected layer with the hyperbolic tangent as activation function, followed by another fully connected layer with the Rectified Linear Unit (ReLU) activation function. The Sequential Click Model consists of an LSTM layer with the hyperbolic tangent activation function, followed by a fully connected layer with the ReLU activation function. Both models are implemented with a 1:1 ratio between the number of units in the two hidden layers, and we use the sigmoid activation function on the output layer as described in section \ref{sec:engineered model} and \ref{sec:sequential model}.
\\\\
The goal is to investigate how suitable click data is in the task of predicting purchase intent. We do that by comparing with a model based on some other data. A \textit{demography} model is chosen for this comparison. We base the model on demographic features of the user. Since the prediction target is immediate purchase intent, we supplement the demographic features with features of time and place to distinguish users who appear in multiple sessions having different intentions. The features are presented in table \ref{tab:demographic}. To make a fair comparison, we model the demographic data with a neural network similar to those used for the click data. Since the demographic data is not sequential data, we use an FFNN like the one used for the Engineered Click Model. 

\pdfbookmark[1]{Baselines}{baselines}
\section{Baselines}\label{sec:baselines}
To better infer about model performance, we compare the models against these simple baseline models: (1) a \textit{most-frequent} model that predicts the most frequent label in the training set for all; (2) a \textit{stratified} model that predicts the two classes with the training set's class distribution; (3) a \textit{length} model based solely on session length.

\pdfbookmark[1]{Data Preparation}{preparation}
\section{Data Preparation}\label{sec:preparation}
Even with a neural network’s powerful representation ability, getting a good quality, clean data set is paramount. The experiment works on two different data sets which we prepare separately as follows. 
\pdfbookmark[1]{Click Data}{click}
\subsection{Click Data}\label{subsec:click}
\begin{figure}[]
\centering
\includegraphics[scale=0.65]{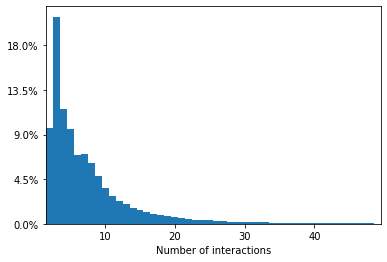}
\caption{Distribution of interactions per session.}
\label{fig:interactions}
\end{figure}

The click data set contains three types of features: interactions with the web site in the form of web page URLs, timestamps of those interactions and interactions within a web page in the form of clicks. The data needs some cleaning before it can be used as input in the click models. Later on we are going to split the data into training, validation and test sets. The following data preprocessing is based solely on the training set.\\
The number of interactions with the website in a session constitutes the session length. The distribution of session lengths in the training set is illustrated in figure \ref{fig:interactions}.
To deal with variable length inputs in the Sequential Click Model, we pad or truncate all sessions to the same length. 99\% of the sessions have length shorter than 30, thus the common session length is set to 30. 
The long sessions are truncated, and the short sessions are padded with zeroes, which is a conventional choice for padding \citep{Dwarampudi2019}.\\
There are 1,064 different web page URLs  in the web page feature, leading to 1,064 categories. Some of the categories have a very sparse occurrence in the training set, which may affect model performance by overfitting. For that reason, we apply binning at categories with too low occurrence. Some of the web page URLs correspond to the same web page, and the difference relies on the path. Thus web page URLs occurring in less than 1\% of the sessions are grouped by their corresponding web page. After binning the web page feature ends up with 47 categories.\\
There are 71 different clicks in the click feature, leading to 71 categories. There is no natural way to group these clicks, thus we remove categories with too low occurrence (frequency less than 1\%) in the click feature. The feature ends up with 20 categories. \\
We transform the timestamp feature to record time between interactions, as we assume this to have greater predictive power than the specific point in time.\\
Finally, we convert the categorical web page and click features with one hot encoding, since their categories do not have a natural ordered relationship. In order to obtain a comparable scale among features, we standardize the numeric time feature to have a mean of 0 and a standard deviation of 1. \\
The Sequential Click Model works directly on this data. For the Engineered Click Model, we extract the features presented in section \ref{subsec:engineering} from this data. 

\pdfbookmark[1]{Demographic Data}{demographic}
\subsection{Demographic Data}\label{subsec:demographic}
The demographic data contains the features presented in table \ref{tab:demographic}. As previously mentioned, we supplement the demographic features of the user with features of time and place. The time features describe when the session took place, while the place features describe where and in what context the session took place. 
\begin{table}[]
\centering
\subfloat[][Demographic features of the user.]{
\scalebox{0.85}{
\begin{tabular}{@{}p{4cm}>{\raggedright}p{6cm}>{\raggedright}p{3cm}>{\raggedleft\arraybackslash}p{2cm}@{}}
\toprule
Feature & Description & Type & \multicolumn{1}{l}{Coverage} \\ \midrule
\rowcolor[HTML]{EFEFEF} 
Age & Measured in whole years. & Numeric & 100.00\% \\
Gender & Man or woman. & Categorical & 100.00\% \\
\rowcolor[HTML]{EFEFEF} 
Income Class & Four intervals. & Ordinal & 99.48\% \\
Education Level & Five levels. & Ordinal & 99.48\% \\
\rowcolor[HTML]{EFEFEF} 
Marital Status & Single or couple. & Categorical & 99.48\% \\
Property Type & \begin{tabular}[c]{@{}l@{}}Rented home, owned \\ home or equity sharing.\end{tabular} & Categorical & 98.16\% \\
\rowcolor[HTML]{EFEFEF} 
Geographic Region & Region of Denmark. & Categorical & 99.67\% \\
Urban Density & \begin{tabular}[c]{@{}l@{}}Metropolis, province, \\ village or countryside.\end{tabular} & Categorical & 99.49\% \\
\rowcolor[HTML]{EFEFEF} 
Children & None, one, two, more than two. & Ordinal & 99.48\% \\
Employment & \begin{tabular}[c]{@{}l@{}}Worker, retired, student \\ or unemployed.\end{tabular} & Categorical & 99.48\% \\ \bottomrule
\end{tabular}}}

\subfloat[][Time features of the session.]{
\scalebox{0.85}{
\begin{tabular}{@{}p{4cm}>{\raggedright}p{6cm}>{\raggedright}p{3cm}>{\raggedleft\arraybackslash}p{2cm}@{}}
\toprule
Feature & Description & Type & \multicolumn{1}{l}{Coverage} \\ \midrule
\rowcolor[HTML]{EFEFEF} 
Month & January to December. & Cyclic & 100.00\% \\
Time Of The Month & \begin{tabular}[c]{@{}l@{}}The beginning, the middle \\ or the end of the month.\end{tabular} & Cyclic & 100.00\% \\
\rowcolor[HTML]{EFEFEF} 
Weekday & Monday to Sunday. & Cyclic & 100.00\% \\
Time Of The Day & \begin{tabular}[c]{@{}l@{}}Morning, forenoon, noon, \\ afternoon, evening or night.\end{tabular} & Cyclic & 100.00\% \\ \bottomrule
\end{tabular}}}

\subfloat[][Place features of the session.]{
\scalebox{0.85}{
\begin{tabular}{@{}p{4cm}>{\raggedright}p{6cm}>{\raggedright}p{3cm}>{\raggedleft\arraybackslash}p{2cm}@{}}
\toprule
Feature & Description & Type & \multicolumn{1}{l}{Coverage} \\ \midrule
\rowcolor[HTML]{EFEFEF} 
Location & \begin{tabular}[c]{@{}l@{}}Home district, neighboring \\ district or foreign district.\end{tabular} & Categorical & 100.00\% \\
Operating System & \begin{tabular}[c]{@{}l@{}}Windows, macOS, \\ Android, iOS or other.\end{tabular} & Categorical & 100.00\% \\
\rowcolor[HTML]{EFEFEF} 
Previous Visits & Number of previous sessions. & Numeric & 100.00\% \\
Distance To Last Visit & Time since last session. & Numeric & 44.00\% \\
\rowcolor[HTML]{EFEFEF} 
Browser & \begin{tabular}[c]{@{}l@{}}Google, Apple, Microsoft, \\ Mozilla or other.\end{tabular} & Categorical & 100.00\% \\ \bottomrule
\end{tabular}}}

\caption{Overview of the demographic features.}
\label{tab:demographic}
\end{table}
As for the click data, the choices we make during data preprocessing are based on the training set.\\
Unlike the click data, the demographic data does not suffer from sparsity in any of the features, in turn the demographic data contains some missing values. Table \ref{tab:demographic} provides an overview of the coverage of non-missing data and the different feature types. We handle the problem with some simple data imputation. In the numeric features we replace missing values with the average of the feature, in the ordinal features we replace missing values with the median of the feature, and in the categorical features we replace missing values with the most frequent category of the feature. Missing values in the feature called "Distance To Last Visit" are treated specially. Since a missing value in this feature is due to the fact that no previous visits have occurred, we replace missing values with the maximum of the feature. \\
We one hot encode the categorical features, and we encode the ordinal features with integers preserving the ordinal relationship of the feature. Initially, we integer encode the cyclic features as well, then we transform them into two dimensions using a sine and cosine transformation to preserve the cyclic relationships. For a cyclic feature $x$ we use the following transformations:
\begin{align}
    x_{sin} &= \sin\left(\frac{2 \pi x}{\max(x)}\right) \\
    x_{cos} &= \cos\left(\frac{2 \pi x}{\max(x)}\right).
\end{align}
Finally, we standardize the numeric, ordinal and cyclic features to have a mean of 0 and a standard deviation of 1.

\pdfbookmark[1]{Evaluation Measures}{measures}
\section{Evaluation Measures}\label{sec:measures}
We evaluate the models by their effectiveness at predicting purchase intent using some standard classification evaluation metrics. Because of the class imbalance in the data, the commonly used performance measure Accuracy is highly misleading, since predicting all observations as the most frequent class will yield a high Accuracy. Instead, we use the Balanced Accuracy, which is given by the following formula \citep{Brodersen2010}:
\begin{align}
    \text{Balanced Accuracy} = \frac{\text{TP}}{2N^+}+\frac{\text{TN}}{2N^-},
\end{align}
where TP (True Positives) is the number of observations correctly predicted as positives, TN (True Negatives) is the number of observations correctly predicted as negatives, $N^+$ and $N^-$ are the number of observations in the actual positive and negative class, respectively. The Balanced Accuracy overcomes the problem, since predicting everything as the most frequent class will yield a Balanced Accuracy of 0.5. However, predicting everything as the least frequent class will also yield a Balanced Accuracy of 0.5, so this metric should not stand alone. Thus we supplement the Balanced Accuracy with the Precision and Recall metrics. They are defined as
\begin{align}
    \text{Precision} &= \frac{\text{TP}}{\text{TP}+\text{FN}} \\
    \text{Recall} &= \frac{\text{TP}}{\text{TP}+\text{FP}}~,
\end{align}
where FN (False Negatives) is the number of observations wrongly predicted as negatives, and FP (False Positives) is the number of observations wrongly predicted as positives. The Recall can trivially be improved by predicting all observations as positive, but the Precision will then suffer. \\
The models to be evaluated are all models that assign to each observation the probability of belonging to the positive class. As a consequence, the metrics presented above all depend on the choice of threshold to separate the two classes from the assigned probabilities. 
The generally used classification threshold of 0.5 is usually unsuitable for an imbalanced classification. Therefore, we use a threshold that separates the two classes with regard to the class distribution in the data set (excluding the test set). That is, for every model we choose the thresholds that predicts 13.12\% as positives. \\
We will also evaluate the models with a metric that takes into account the different threshold values. The Area Under the ROC Curve (AUC) is used for this purpose. Let $\theta$ be a parameter denoting the threshold. The AUC is the area measured under the curve that appears, when plotting the True Positive Rate (TPR) against the False Positive Rate (FPR) for all conceivable values of $\theta$. The TPR and FPR as functions of $\theta$ are computed as
\begin{align}
    \text{FPR}(\theta) &= \frac{\text{FP}(\theta)}{\text{FP}(\theta)+\text{TN}(\theta)} \\
    \text{TPR}(\theta) &= \frac{\text{TP}(\theta)}{\text{TP}(\theta)+\text{FN}(\theta)}.
\end{align}
Besides the desirable property of not depending on the choice of threshold, the AUC measure also accounts for imbalanced classes due to the normalization of the TPR and FPR by the number of observations in the true and false class respectively.

\pdfbookmark[1]{Tuning}{tuning}
\section{Tuning}\label{sec:tuning}
We split the sample of user sessions into training, validation and test sets. The split ratio is 80\% for training, 10\% for validation and 10\% for test. We use out-of-time validation and testing, since the purpose of the model is to predict ahead of time. \\
We fit all the neural models with the Adam optimization algorithm \citep{Goodfellow2016}. The algorithm is an extension to the computationally efficient algorithm, stochastic gradient descent. It is enhanced with adaptive learning rates, making the performance more robust. We apply the Adam optimizer with the default configuration parameters in TensorFlow's implementation. \\
We add regularization to prevent the models from learning the noise in the training data and not performing well on the test data. Dropout is a method to prevent overfitting that has the same powerful effect as ensembles of many neural networks but is less computationally expensive \citep{Goodfellow2016}. Thus we add dropout to each model after the first hidden layer.
\\
We use hyperparameter tuning to choose batch size, number of hidden units and dropout rate. We use early stopping to determine an efficient number of epochs during training. We measure model performance on the validation data with AUC. The main grid search results are presented in table \ref{tab:Grid}. We test powers of 2 for the batch size and number of units in the hidden layer to offer better run time when using GPUs. We test the dropout rate for a range between 0 and 1, as it is a probability. We test all the values in [0,1] with step 0.1, and the best results are reported in table \ref{tab:Grid}.

\begin{table}[]
\centering
\subfloat[][Demography Model.]{
\begin{tabular}{@{}lrrrrrrrrr@{}}
\toprule
           & \multicolumn{3}{c}{Dropout 0.2}                                              & \multicolumn{3}{c}{Dropout 0.3}                                              & \multicolumn{3}{c}{Dropout 0.4}                                           \\ \midrule
Batch size & \multicolumn{1}{c}{32} & \multicolumn{1}{c}{64} & \multicolumn{1}{c|}{128}   & \multicolumn{1}{c}{32} & \multicolumn{1}{c}{64} & \multicolumn{1}{c|}{128}   & \multicolumn{1}{c}{32} & \multicolumn{1}{c}{64} & \multicolumn{1}{c}{128} \\ \midrule
32 units   & 0.727                  & 0.728                  & \multicolumn{1}{r|}{0.729} & 0.729                  & \textbf{0.731}         & \multicolumn{1}{r|}{0.727} & 0.721                  & 0.720                  & 0.724                   \\
64 units   & 0.726                  & 0.726                  & \multicolumn{1}{r|}{0.727} & 0.726                  & 0.729                  & \multicolumn{1}{r|}{0.721} & 0.727                  & 0.729                  & 0.729                   \\
128 units  & 0.725                  & 0.727                  & \multicolumn{1}{r|}{0.727} & 0.721                  & 0.728                  & \multicolumn{1}{r|}{0.725} & 0.720                  & 0.729                  & 0.728                   \\ \bottomrule
\end{tabular}}

\subfloat[][Engineered Click Model.]{
\begin{tabular}{@{}lrrrrrrrrr@{}}
\toprule
           & \multicolumn{3}{c}{Dropout 0.2}                                              & \multicolumn{3}{c}{Dropout 0.3}                                              & \multicolumn{3}{c}{Dropout 0.4}                                           \\ \midrule
Batch size & \multicolumn{1}{c}{32} & \multicolumn{1}{c}{64} & \multicolumn{1}{c|}{128}   & \multicolumn{1}{c}{32} & \multicolumn{1}{c}{64} & \multicolumn{1}{c|}{128}   & \multicolumn{1}{c}{32} & \multicolumn{1}{c}{64} & \multicolumn{1}{c}{128} \\ \midrule
32 units   & 0.766                  & 0.770                  & \multicolumn{1}{r|}{0.770} & 0.764                  & 0.764                  & \multicolumn{1}{r|}{0.766} & 0.760                  & 0.762                  & 0.762                   \\
64 units   & 0.774                  & 0.767                  & \multicolumn{1}{r|}{0.766} & 0.763                  & 0.770                  & \multicolumn{1}{r|}{0.771} & 0.766                  & 0.770                  & 0.773                   \\
128 units  & 0.774                  & 0.774                  & \multicolumn{1}{r|}{0.775} & 0.768                  & \textbf{0.777}         & \multicolumn{1}{r|}{0.772} & 0.768                  & 0.768                  & 0.776                   \\ \bottomrule
\end{tabular}}

\subfloat[][Sequential Click Model.]{
\begin{tabular}{@{}lrrrllllll@{}}
\toprule
           & \multicolumn{3}{c}{Dropout 0.3}                                              & \multicolumn{3}{c}{Dropout 0.4}                                                       & \multicolumn{3}{c}{Dropout 0.5}                                           \\ \midrule
Batch size & \multicolumn{1}{c}{32} & \multicolumn{1}{c}{64} & \multicolumn{1}{c|}{128}   & \multicolumn{1}{c}{32} & \multicolumn{1}{c}{64} & \multicolumn{1}{c|}{128}            & \multicolumn{1}{c}{32} & \multicolumn{1}{c}{64} & \multicolumn{1}{c}{128} \\ \midrule
32 units   & 0.812                  & 0.810                  & \multicolumn{1}{r|}{0.811} & 0.811                  & 0.810                  & \multicolumn{1}{l|}{0.810}          & 0.813                  & 0.811                  & 0.810                   \\
64 units   & 0.812                  & 0.811                  & \multicolumn{1}{r|}{0.811} & 0.811                  & 0.812                  & \multicolumn{1}{l|}{\textbf{0.814}} & 0.813                  & 0.811                  & 0.813                   \\
128 units  & 0.810                  & 0.813                  & \multicolumn{1}{r|}{0.810} & 0.811                  & 0.812                  & \multicolumn{1}{l|}{0.812}          & 0.811                  & 0.812                  & 0.811                   \\ \bottomrule
\end{tabular}}
\caption{Grid search results for dropout rate, number of units and batch size, evaluated with AUC.}
\label{tab:Grid}
\end{table}



\pdfbookmark[1]{Results}{results}
\section{Results}\label{sec:results}
\begin{figure}[]
\centering
\includegraphics[scale=0.75]{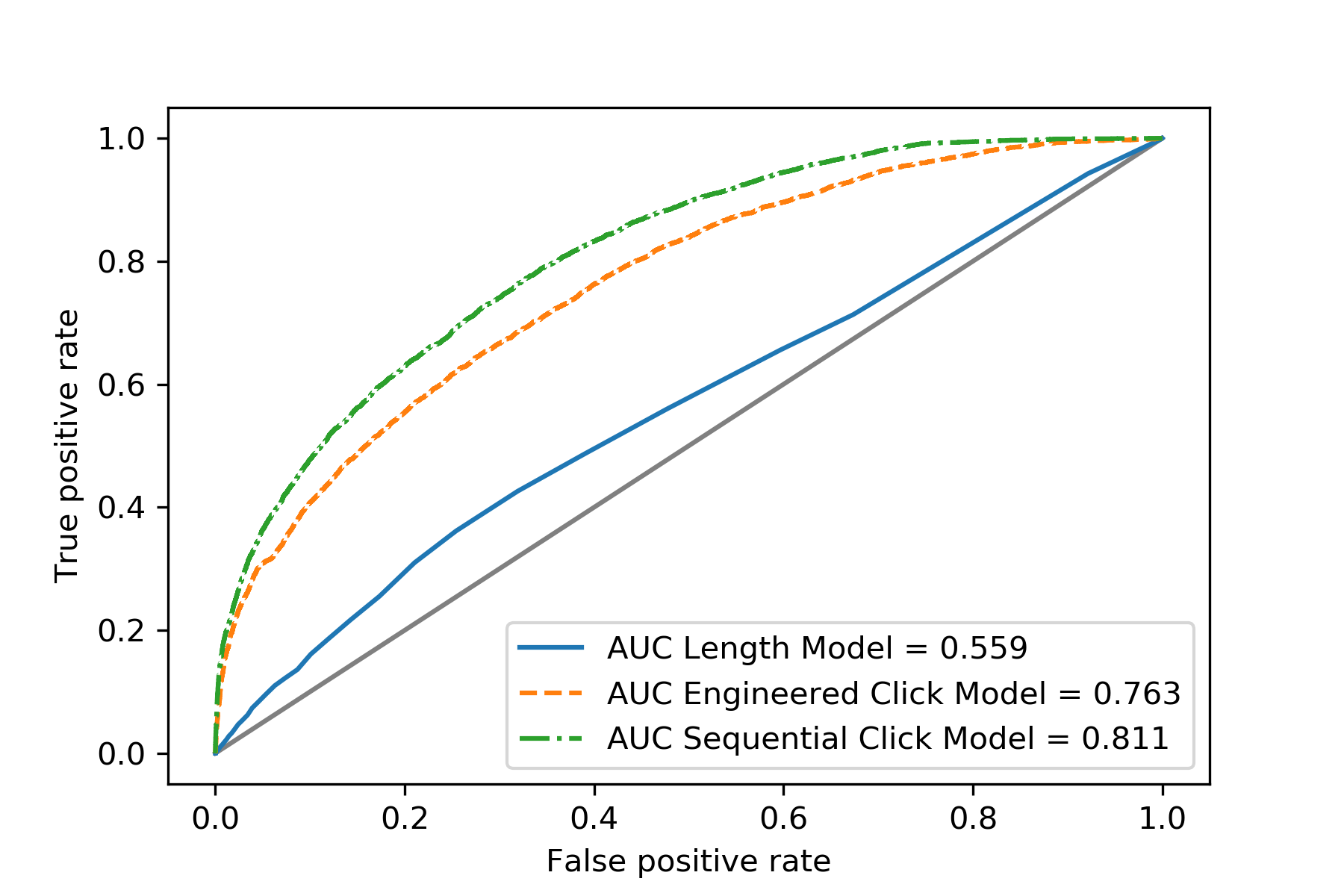}
\caption{ROC Curves of the two click-based models compared to the Length Model.}
\label{fig:ROC1}
\end{figure}

\begin{figure}[]
\centering
\includegraphics[scale=0.75]{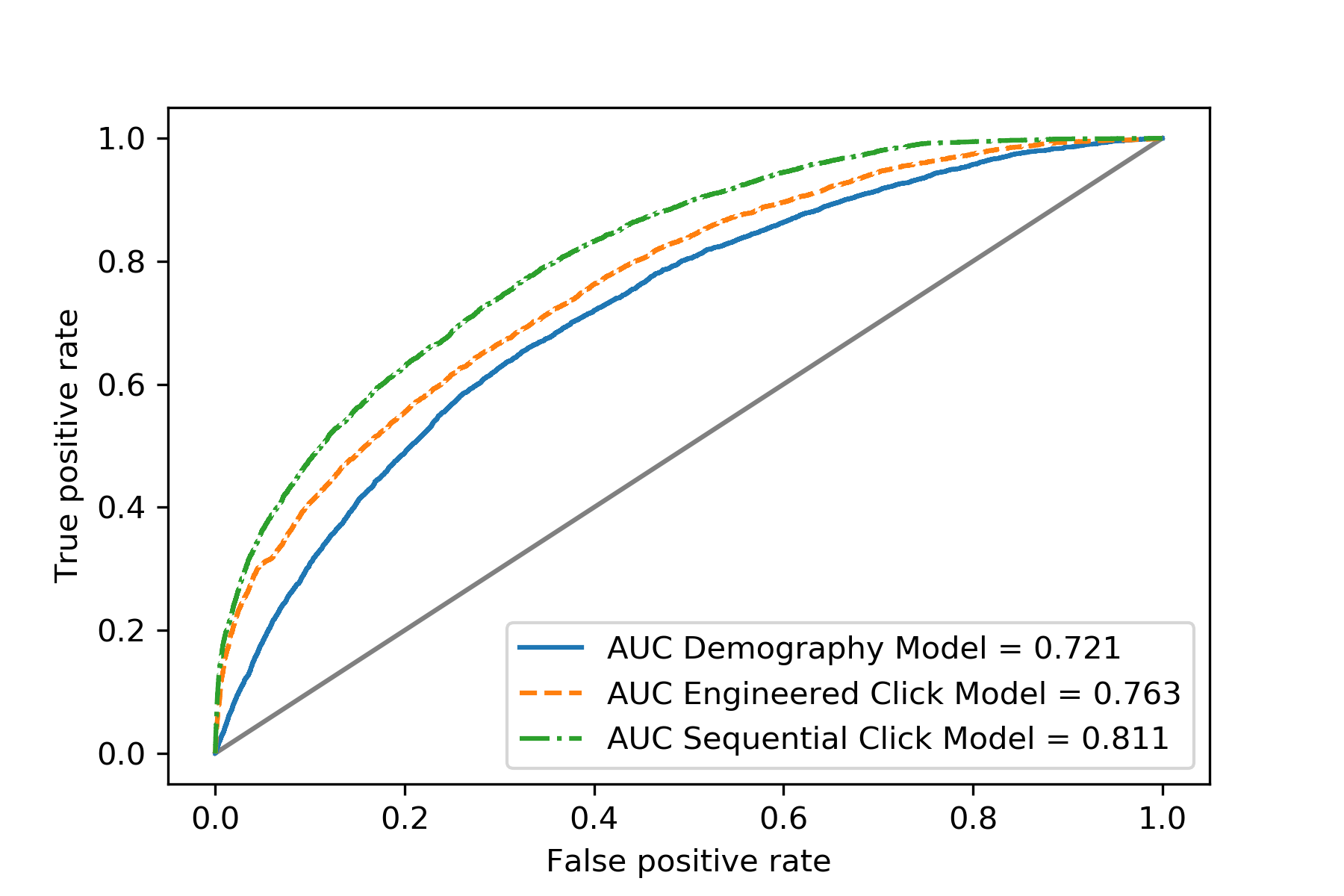}
\caption{ROC Curves of the two click-based models compared to the Demography Model.}
\label{fig:ROC2}
\end{figure}

Table \ref{tab:evaluation} summarizes performance on the test set of all the models conducted.
\begin{table}[]
\centering
\begin{tabular}{@{}lrrrr@{}}
\toprule
\multicolumn{1}{c}{}   & \multicolumn{1}{c}{Balanced Accuracy} & \multicolumn{1}{c}{Precision} & \multicolumn{1}{c}{Recall} & \multicolumn{1}{c}{AUC} \\ \midrule
Most-frequent Model    & 0.5000                                & 0.0000                        & -                          & -                       \\
Stratified Model       & 0.4983                                & 0.1213                        & 0.1194                     & 0.4983                  \\
Length Model           & 0.5332                                & 0.1812                        & 0.1853                     & 0.5593                  \\ \midrule
Demography Model       & 0.6412                                & 0.2636                        & 0.4680                     & 0.7207                  \\ \midrule
Engineered Click Model & 0.6726                                & 0.3005                        & 0.5157                     & 0.7634                  \\
Sequential Click Model & \textbf{0.7011}                                & \textbf{0.3648}                        & \textbf{0.5343}                     & \textbf{0.8109}                  \\ \bottomrule
\end{tabular}
\caption{All models are evaluated with some standard classification evaluation metrics on the test set. Note that it is not possible to compute the Recall score for the Most-frequent Model as the denominator is zero, nor is it possible to compute AUC as the TPR is zero for all $\theta$.}
\label{tab:evaluation}
\end{table}
Figure \ref{fig:ROC1} illustrates the ROC Curves of the two click-based models.
It appears that both click models are considerably better than the the Length Model and the diagonal line, representing a random guess. Moreover, the Sequential Click Model outperforms the Engineered Click Model. Compared to the Length Model, the Engineered Click Model increases AUC with 36.5\%, and the Sequential Click Model increases AUC with 45\%. The metrics presented in table \ref{tab:evaluation} are all consistent with the results of the ROC Curves, with the click models in particular improving the Recall score relative to the baseline models. The Sequential Click model outperforms the Engineered Click Model across all metrics in table \ref{tab:evaluation} as well. \\
Figure \ref{fig:ROC2} illustrates the ROC Curves of the two click models compared to the Demography Model.
It appears that both click models outperform the Demography Model, as the Engineered Click Model increases AUC with 5.9\%, and the Sequential Click Model increases AUC with 12.5\% compared to the Demography Model. 
The metrics in table \ref{tab:evaluation} indicate the same, as the click models outperform the Demography Model across all metrics, especially in the Precision score.

\pdfbookmark[1]{Focused Analysis}{analysis}
\section{Focused Analysis}\label{sec:analysis}
\pdfbookmark[1]{Concatenating Models}{concatenating}
\subsection{Concatenating Models}\label{subsec:concatenating}
\begin{figure}%
    \centering
    \subfloat[][Engineered Click Model concatenated\\ with Demography Model]{\includegraphics[scale=0.51]{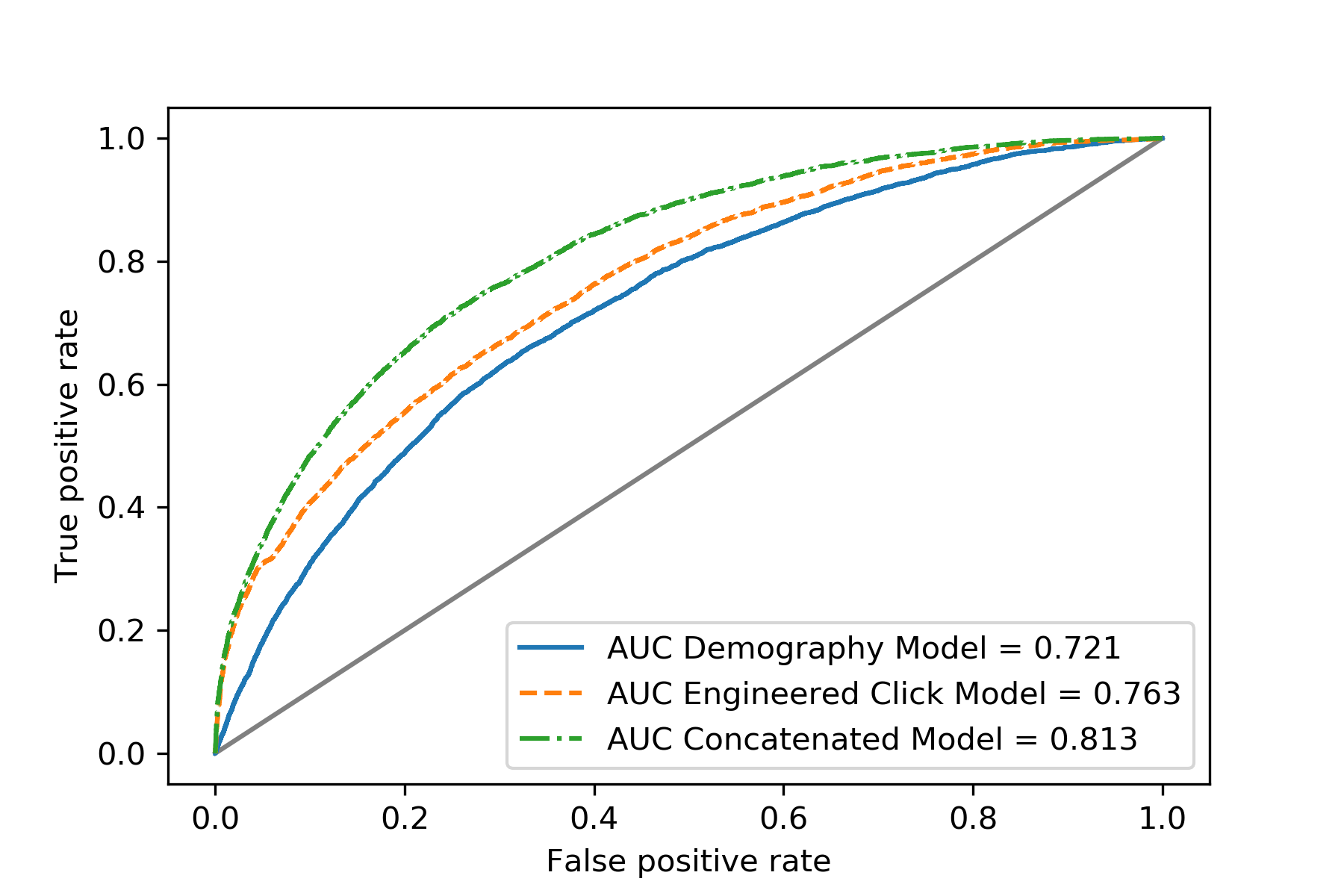}}%
    \subfloat[][Sequential Click Model concatenated\\ with Demography Model]{\includegraphics[scale=0.51]{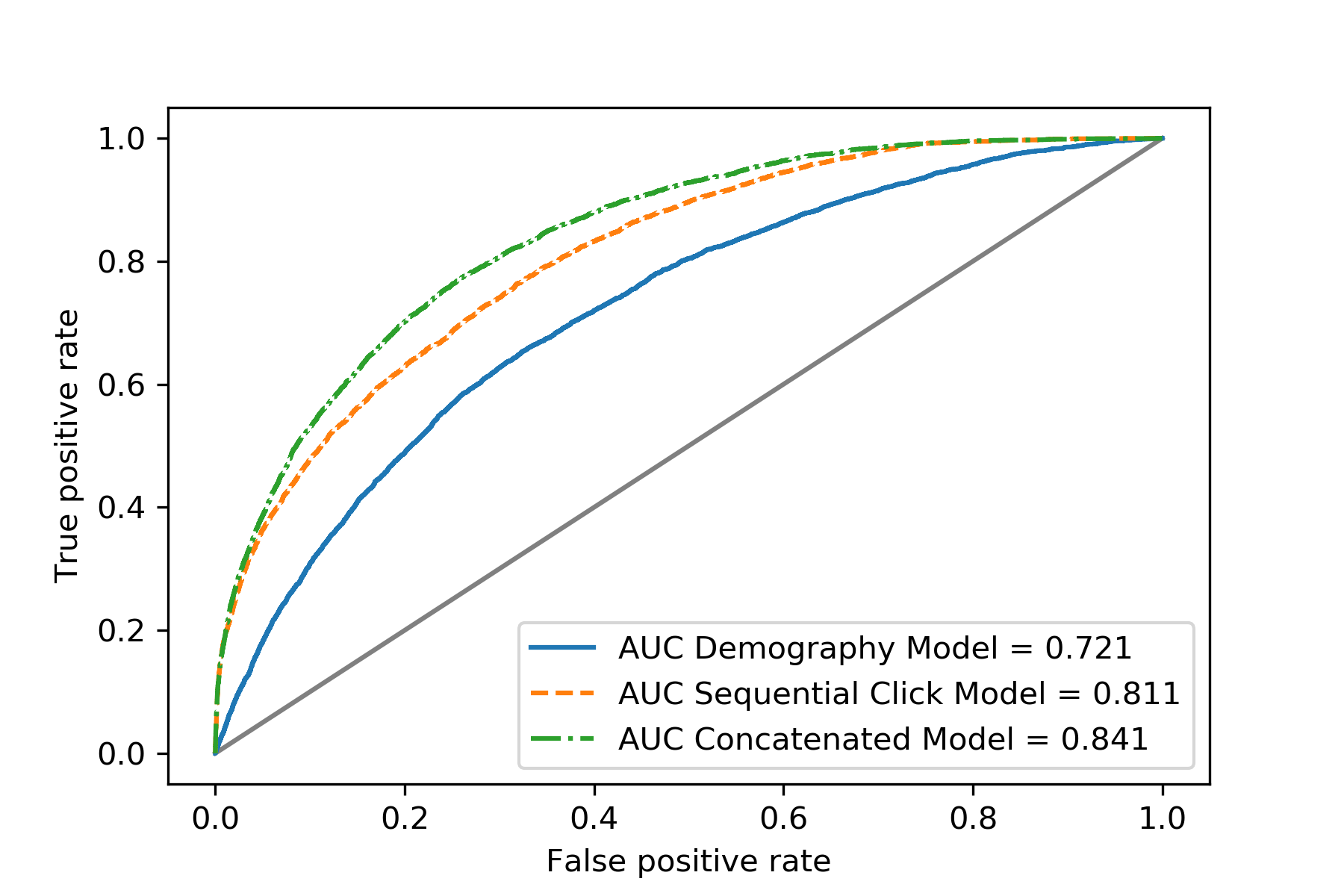}}%
    \caption{ROC Curves of concatenated models.}%
    \label{fig:ROC34}%
\end{figure}

Since both click models outperform the Demography Model, it seems that the click data is most suitable in the task of predicting purchase intent. To further investigate whether the click data captures the same or different aspects of user intentions as the demographic data, we conduct two concatenated models, where the Engineered Click Model and the Sequential Click Model are concatenated with the Demography Model, respectively. The result is illustrated in figure \ref{fig:ROC34}.
It appears that both of the concatenated models yield slightly greater performance than the two click models, suggesting that the two data types mainly captures the same aspects of user intentions, but in combination they are able to catch even new aspects. The reason might be that click data provides context around a user’s intent, while demographic data provides context around a user’s ability to make purchases.
\pdfbookmark[1]{Error Analysis}{error}
\subsection{Error Analysis}\label{subsec:error}
\begin{figure}[]%
    \centering
    \subfloat{
        \raisebox{-.5\height}{\includegraphics[scale=0.58]{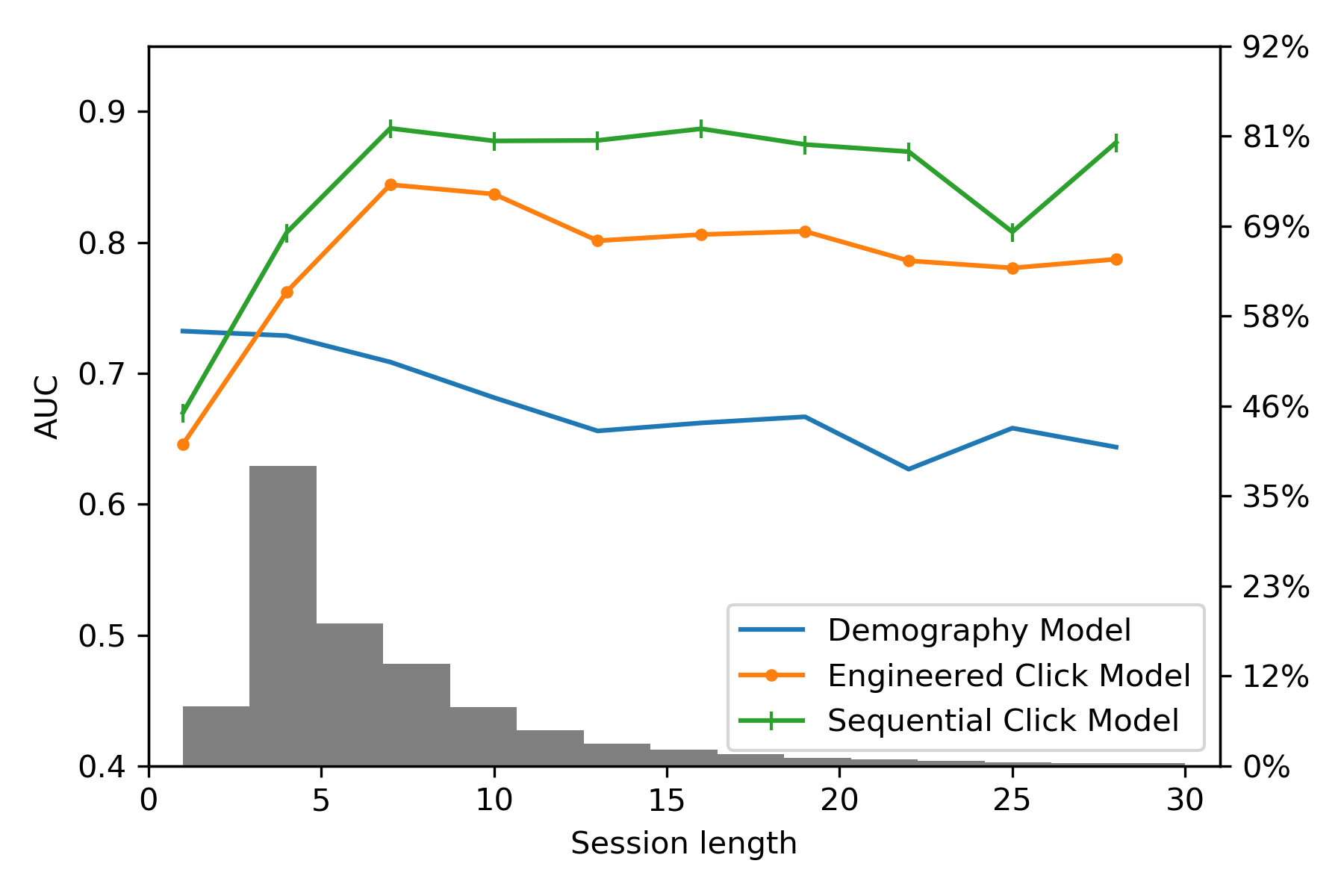}}
    }%
    \subfloat{
         \scalebox{0.58}{
         \begin{tabular}{@{}lrrr@{}}
         \toprule
         \multicolumn{1}{c}{\begin{tabular}[c]{@{}c@{}}Session\\ Length\end{tabular}} & \multicolumn{1}{c}{\begin{tabular}[c]{@{}c@{}}Demography\\ Model\end{tabular}} & \multicolumn{1}{c}{\begin{tabular}[c]{@{}c@{}}Engineered\\ Click Model\end{tabular}} & \multicolumn{1}{c}{\begin{tabular}[c]{@{}c@{}}Sequential\\ Click Model\end{tabular}} \\ \midrule
         1-3 & 0.7322 & 0.6457 & 0.6698 \\
         4-6 & 0.7287 & 0.7622 & 0.8074 \\
         7-9 & 0.7086 & 0.8441 & 0.8872 \\
         10-12 & 0.6814 & 0.8369 & 0.8775 \\
         13-15 & 0.6560 & 0.8013 & 0.8779 \\
         16-18 & 0.6621 & 0.8060 & 0.8868 \\
         19-21 & 0.6668 & 0.8084 & 0.8748 \\
         22-24 & 0.6267 & 0.7860 & 0.8693 \\
         25-27 & 0.6581 & 0.7805 & 0.8080 \\
         28-30 & 0.6436 & 0.7872 & 0.8763 \\ \bottomrule
         \end{tabular}
         }
    }%
    \caption{AUC by session length for the two click-based models and the Demography Model.}%
    \label{fig:Length}%
\end{figure}

\begin{figure}[]%
    \centering
    \subfloat{
        \raisebox{-.5\height}{\includegraphics[scale=0.58]{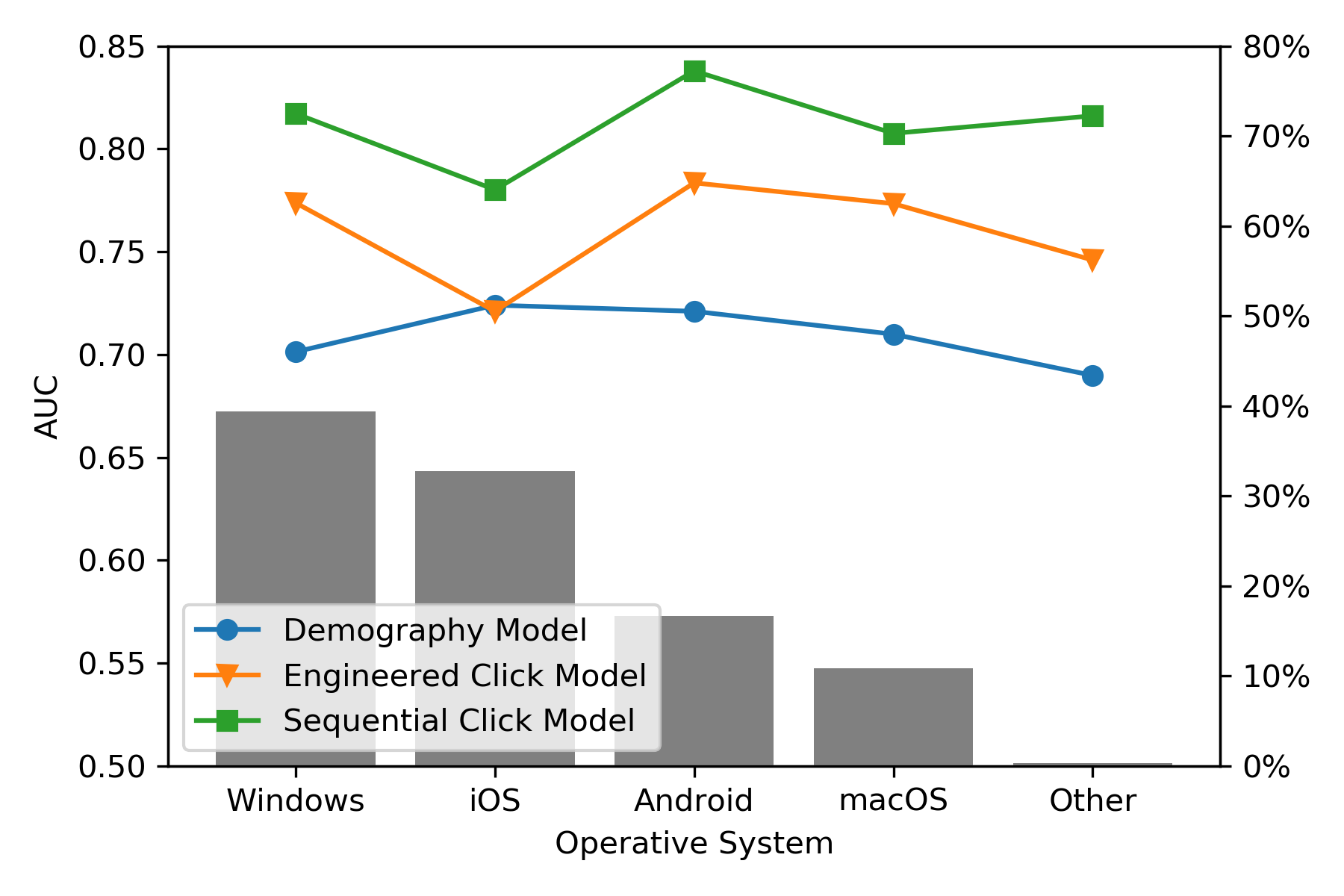}}
    }%
    \subfloat{
         \scalebox{0.58}{
         \begin{tabular}{@{}lrrr@{}}
         \toprule
         \multicolumn{1}{c}{\begin{tabular}[c]{@{}c@{}}Operating\\ System\end{tabular}} & \multicolumn{1}{c}{\begin{tabular}[c]{@{}c@{}}Demography\\ Model\end{tabular}} & \multicolumn{1}{c}{\begin{tabular}[c]{@{}c@{}}Engineered\\ Click Model\end{tabular}} & \multicolumn{1}{c}{\begin{tabular}[c]{@{}c@{}}Sequential\\ Click Model\end{tabular}} \\ \midrule
         Windows & 0.7013 & 0.7738 & 0.8172 \\
         iOS & 0.7240 & 0.7211 & 0.7802 \\
         Android & 0.7210 & 0.7835 & 0.8379 \\
         macOS & 0.7099 & 0.7734 & 0.8075 \\
         Other & 0.6899 & 0.7460 & 0.8161 \\ \bottomrule
         \end{tabular}
         }
    }%
    \caption{AUC by Operating System for the two click-based models and the Demography Model.}%
    \label{fig:OS}%
\end{figure}

The session length represents a useful signal in the test data to analyze models performance. 
In figure \ref{fig:Length} we graph the main three models with AUC score broken down by session length. For context, the distribution of session length in the test set is also provided, and a table with the AUC scores across intervals of session lengths. 
Both click models underperform on sessions of length 1-3, so clearly it is difficult to predict the purchase intent with such a small input signal. From length 7 and upwards, we obtain an average AUC of 0.8318 for the Engineered Click Model and an average AUC of 0.8827 for the Sequential Click model. Furthermore, the difference in model performance between the two click models is smaller for shorter sessions. The performance of the Demography Model does not depend on session length, and the Demography Model outperform both click models on very short sessions. However, there is a slight tendency for the Demography Model to be less accurate the longer the sessions. The reason for this is most likely the finding of purchase intent growing with session length combined with the fact that the Demography Model generally is poor at identifying the positive class, as seen from the Precision and Recall scores.
\\\\
The website from which we collected the data is identical for all Operating Systems. However, since Operating System is closely related to device, it is interesting to consider how the models perform across different Operating Systems. In particular if the click patterns are equally difficult to classify on computer devices compared to tablet devices. The results are presented in figure \ref{fig:OS}. Note that Windows and macOS belong to computer devices, while iOS and Android belong to tablet devices. 
\\
It appears that all three models are quiet robust across the different types of Operating Systems, as there are no major differences in model performance. Moreover, performances of the two click models do not depend on the type of device. In fact the greatest difference occurs between two Operating Systems belonging to the same device, namely iOS and Android.
\pdfbookmark[1]{Temporal Order Analysis}{temporal}
\subsection{Temporal Order Analysis}\label{subsec:temporal}
The Engineered Click Model and the Sequential Click Model are based on the same data, yet there is a difference in the performance of the two models. To better understand the reason for that, we investigate the importance of the temporal order more closely. We do that by random shuffling the temporal order in the data and fit a model on the shuffled data. In an attempt to completely remove the temporal order, we shuffle both training, validation and test data. We perform the experiment five times to take account of randomness. The result is presented in table \ref{tab:temporal}.
\begin{table}[]
\centering
\begin{tabular}{@{}lrrrrr@{}}
\toprule
                 & \multicolumn{1}{c}{AUC} & \multicolumn{1}{c}{\begin{tabular}[c]{@{}c@{}}Relative change\\ in AUC\end{tabular}} & \multicolumn{1}{c}{\begin{tabular}[c]{@{}c@{}}Balanced\\ Accuracy\end{tabular}} & \multicolumn{1}{c}{Precision} & \multicolumn{1}{c}{Recall} \\ \midrule
Original order   & 0.8109                  &                                                                                      & 0.7011                                                                          & 0.3648                        & 0.5343                     \\ \midrule
Random shuffle 1 & 0.7797                  & -3.85\%                                                                              & 0.6821                                                                          & 0.3193                        & 0.5224                     \\
Random shuffle 2 & 0.7798                  & -3.84\%                                                                              & 0.6864                                                                          & 0.3064                        & 0.5493                     \\
Random shuffle 3 & 0.7800                  & -3.81\%                                                                              & 0.6786                                                                          & 0.3347                        & 0.4975                     \\
Random shuffle 4 & 0.7816                  & -3.61\%                                                                              & 0.6791                                                                          & 0.3468                        & 0.4890                     \\
Random shuffle 5 & 0.7774                  & -4.13\%                                                                              & 0.6836                                                                          & 0.3161                        & 0.5300                     \\ \midrule
Mean             & 0.7797                  & -3.85\%                                                                              & 0.6820                                                                          & 0.3247                        & 0.5176                     \\ \bottomrule
\end{tabular}
\caption{Result of the Sequential Click Model after random shuffling the temporal order of the whole data set.}
\label{tab:temporal}
\end{table}
As expected the temporal order seems to have an impact, but the Sequential Click Model still performs better than the Engineered Click Model. Besides the ability to incorporate the temporal order, it suggests that the Sequential Click Model also benefits from the ability to do automatic feature learning instead of manual feature engineering as used for the Engineered Click Model.
\pdfbookmark[1]{Feature Ablation Study}{ablation}
\subsection{Feature Ablation Study}\label{subsec:ablation}
We perform a feature ablation study to understand the contribution of features to the performance of the main models. The result is presented in table \ref{tab:ablation}.
\begin{table}[]
\centering
\subfloat[][Demography Model.]{
\begin{tabular}{@{}lrrrrr@{}}
\toprule
 & \multicolumn{1}{c}{AUC} & \multicolumn{1}{c}{\begin{tabular}[c]{@{}c@{}}Relative change\\ in AUC\end{tabular}} & \multicolumn{1}{c}{\begin{tabular}[c]{@{}c@{}}Balanced\\ Accuracy\end{tabular}} & \multicolumn{1}{c}{Precision} & \multicolumn{1}{c}{Recall} \\ \midrule
All features & 0.7207 &  & 0.6412 & 0.2636 & 0.4680 \\ \midrule
No demographic features & 0.7025 & -2.52\% & 0.6251 & 0.2605 & 0.4191 \\
No time features & 0.7102 & -1.46\% & 0.6275 & 0.2563 & 0.4338 \\
No place features & 0.6609 & -8.30\% & 0.5890 & 0.2421 & 0.3205 \\ \bottomrule
\end{tabular}}

\subfloat[][Engineered Click Model.]{
\begin{tabular}{@{}lrrrrr@{}}
\toprule
 & \multicolumn{1}{c}{AUC} & \multicolumn{1}{c}{\begin{tabular}[c]{@{}c@{}}Relative change\\ in AUC\end{tabular}} & \multicolumn{1}{c}{\begin{tabular}[c]{@{}c@{}}Balanced\\ Accuracy\end{tabular}} & \multicolumn{1}{c}{Precision} & \multicolumn{1}{c}{Recall} \\ \midrule
All features & 0.7634 &  & 0.6726 & 0.3005 & 0.5157 \\ \midrule
No web page features & 0.7542 & -1.21\% & 0.6710 & 0.3000 & 0.5045 \\
No click features & 0.7376 & -3.38\% & 0.6408 & 0.2911 & 0.4303 \\
No time features & 0.7504 & -1.71\% & 0.6640 & 0.3145 & 0.4750 \\ \bottomrule
\end{tabular}}

\subfloat[][Sequential Click Model.]{
\begin{tabular}{@{}lrrrrr@{}}
\toprule
 & \multicolumn{1}{c}{AUC} & \multicolumn{1}{c}{\begin{tabular}[c]{@{}c@{}}Relative change\\ in AUC\end{tabular}} & \multicolumn{1}{c}{\begin{tabular}[c]{@{}c@{}}Balanced\\ Accuracy\end{tabular}} & \multicolumn{1}{c}{Precision} & \multicolumn{1}{c}{Recall} \\ \midrule
All features & 0.8109 &  & 0.7011 & 0.3648 & 0.5343 \\ \midrule
No web page feature & 0.7848 & -3.21\% & 0.6761 & 0.3318 & 0.4932 \\
No click feature & 0.7954 & -1.91\% & 0.6881 & 0.3593 & 0.5036 \\
No time feature & 0.7952 & -1.94\% & 0.6912 & 0.3547 & 0.5157 \\ \bottomrule
\end{tabular}}
\caption{Feature ablation showing the model performance when removing different groups of features.}
\label{tab:ablation}
\end{table}
It appears that no features are redundant, as model performance decreases in all of the experiments.\\
For the Demography Model, we have expected that the demographic features cannot stand alone. It appears that the place features contribute a lot in conjunction with the demographic features, while the time features contribute a lot less. \\
It is interesting how the web page feature gives rise to the biggest drop in performance of the Sequential Click Model, while this is the case for the click features in the Engineered Click Model. The web page feature is fundamental in the Sequential Click Model, most likely because it also provides context for the time feature. By removing the web page feature, the connection between time and on which web page the user spent that amount of time is lost. This connection is incorporated into the time features in the Engineered Click Model, since we constructed the time features by computing the time spent on each web page.\\
The fact that the click features contribute a lot to the Engineered Click Model and less to the Sequential Click Model, is either because the Sequential Click Model has a better ability to perform with only the web page feature and the time feature, or because the feature engineering outperforms the feature learning (in the Sequential Click Model) of the click feature. 
\pdfbookmark[1]{Imbalanced Data}{imbalance}
\subsection{Imbalanced Data}\label{subsec:imbalance}
In section \ref{sec:measures} we discussed how some evaluation metrics, such as Accuracy, incorrectly indicate good performance. Besides this problem, learning from imbalanced data sets can also be very difficult.
When class imbalance exists within training data, learners will typically overclassify the majority class due to its increased prior probability. As a result, the observations belonging to the minority class are misclassified more often than those belonging to the majority class \citep{Johnson2019}.\\
A number of methods for addressing class imbalance have been proposed. That includes data-level methods that work across different machine learning models and algorithm-level methods that are linked to specific models.\\
None of the models conducted in this thesis addresses the problem of imbalanced data. It is not crucial since the aim is to compare model performance rather than achieving high performance, and since the level of class imbalance is the same for all models. However, different data types, different data representations and different network architectures will react differently when addressing the problem of imbalanced data, leading to a better comparison. To this end we perform an experiment to explore the impact of class imbalance on the three main models. We use the two simple techniques, random undersampling (RUS) and random oversampling (ROS), both modifying the training distributions in order to decrease the level of imbalance. RUS discards random observations from the majority class, while ROS duplicates random observations from the minority class. In both cases we resample the training set until a balanced class distribution is obtained with 50 \% in each class. Table \ref{tab:Imbalance} presents the results.
\begin{table}[]
\centering
\subfloat[][Demography Model.]{
\begin{tabular}{@{}lrrrrr@{}}
\toprule
                & \multicolumn{1}{c}{AUC} & \multicolumn{1}{c}{\begin{tabular}[c]{@{}c@{}}Relative change\\ in AUC\end{tabular}} & \multicolumn{1}{c}{\begin{tabular}[c]{@{}c@{}}Balanced\\ Accuracy\end{tabular}} & \multicolumn{1}{c}{Precision} & \multicolumn{1}{c}{Recall} \\ \midrule
Imbalanced data & 0.7207                  &                                                                                      & 0.6412                                                                          & 0.2636                        & 0.4680                     \\ \midrule
RUS             & 0.7127                  & -1.11\%                                                                              & 0.6587                                                                          & 0.2114                        & 0.6745                     \\
ROS             & 0.7027                  & -2.49\%                                                                              & 0.6491                                                                          & 0.2031                        & 0.6735                     \\ \bottomrule
\end{tabular}}

\subfloat[][Engineered Click Model.]{
\begin{tabular}{@{}lrrrrr@{}}
\toprule
                & \multicolumn{1}{c}{AUC} & \multicolumn{1}{c}{\begin{tabular}[c]{@{}c@{}}Relative change\\ in AUC\end{tabular}} & \multicolumn{1}{c}{\begin{tabular}[c]{@{}c@{}}Balanced\\ Accuracy\end{tabular}} & \multicolumn{1}{c}{Precision} & \multicolumn{1}{c}{Recall} \\ \midrule
Imbalanced data & 0.7634                  &                                                                                      & 0.6726                                                                          & 0.3005                        & 0.5157                     \\ \midrule
RUS             & 0.7610                  & -0.31\%                                                                              & 0.6803                                                                          & 0.2572                        & 0.6113                     \\
ROS             & 0.7693                  & 0.78\%                                                                               & 0.6879                                                                          & 0.2447                        & 0.6689                     \\ \bottomrule
\end{tabular}}

\subfloat[][Sequential Click Model.]{
\begin{tabular}{@{}lrrrrr@{}}
\toprule
                & \multicolumn{1}{c}{AUC} & \multicolumn{1}{c}{\begin{tabular}[c]{@{}c@{}}Relative change\\ in AUC\end{tabular}} & \multicolumn{1}{c}{\begin{tabular}[c]{@{}c@{}}Balanced\\ Accuracy\end{tabular}} & \multicolumn{1}{c}{Precision} & \multicolumn{1}{c}{Recall} \\ \midrule
Imbalanced data & 0.8109                  &                                                                                      & 0.7011                                                                          & 0.3648                        & 0.5343                     \\ \midrule
RUS             & 0.7837                  & -3.35\%                                                                              & 0.6963                                                                          & 0.2574                        & 0.6648                     \\
ROS             & 0.8092                  & -0.20\%                                                                              & 0.7197                                                                          & 0.2779                        & 0.6964                     \\ \bottomrule
\end{tabular}}
\caption{Results of using RUS and ROS for balancing the class distribution.}
\label{tab:Imbalance}
\end{table}
\\
Overall, we do not see big improvements when balancing the training data, leading to the conclusion that the scale of imbalance is not critical enough to negatively affect model performance. Only the Recall Score significantly improves from balancing, meaning the models have a greater proportion of correct predictions among the predicted positives. Furthermore, the three models react roughly the same to this experiment, thus it does not affect our comparison either. \\
On the other hand, some decrease in model performance also appears. In particular, the Demography Model has a notable drop in AUC when using ROS, and the Sequential Click Model has a notable drop in AUC when using RUS. This is most likely due to the problem of ROS to cause overfitting, and RUS that reduces the total amount of information the model has to learn from. A variety of intelligent sampling methods have been developed in an attempt to balance these trade-offs \citep{Johnson2019}. Intelligent RUS methods aim to select majority observations for removal based on their distance from minority observations. Intelligent ROS methods have been developed to produce artificial minority observations by interpolating between existing minority observations and their nearest minority neighbors. The intelligent sampling methods are distance based nearest neighbor methods and therefore do not cope with variable length sequence data. Hence in our case, more advanced algorithm-level methods will be needed to improve over the simple RUS and ROS methods. 

\chapter{Conclusion}\label{chap:conclusion}
\pdfbookmark[1]{Conclusion}{conclusion}
\section{Conclusion}\label{sec:conclusion}
This section summarizes the structured inquiry into the open actuarial mathematics problem of modelling user behaviour using machine learning methods, in order to predict purchase intent of non-life insurance products: The objective of the thesis was to model user-clicks from a service-based business website for the purpose of predicting purchase intent, as well as exploring the potential of click data for this task. \\
We used two feature-based models to model user behaviour: (1) a model where we engineered the user interactions with the website into non-temporal features and then used a Feed Forward Neural Network to learn a mapping between features and the user's intention; (2) a model where we represented user interactions at different time steps by a sequence of features and then used a Recurrent Neural Network with Long Short Term Memory cells to learn a mapping between features and the user's intention. We compared the capability of click data to predict purchase intent with the capability of a model based on demographic features.\\
We evaluated our approach on the prediction task by measuring model performance on unseen test data. Our experimental results show that users’ historical interactions with the website have great predictive power which the feature-based models manage to enumerate. In terms of standard evaluation metrics for classification, both of our proposed models significantly improve over simple baseline models. The ability of a Recurrent Neural Network to maintain the temporal order of user interactions and to avoid manual feature engineering yield slightly better performance than the more simple Feed Forward Neural Network.\\
We found that the click data is more discriminative than the demographic data in the task of predicting purchase intent. That said, click data used in conjunction with demographic data can contribute new dimensions to the task. Through an error analysis we saw that meaningful click data must accumulate through a user session, whereas demographic data is independent of the session length. Though the behavioural indicators in the click data are such strong signals that the click-based models are robust across different devices.

\pdfbookmark[1]{Future Work}{future}
\section{Future Work}\label{sec:future}
\pdfbookmark[1]{Social Networking Data}{social network}
\subsection{Social Networking Data}\label{subsec:social network}
A social network is a network of individuals connected by interpersonal relationships. It can for instance be an online site through which people create and maintain interpersonal relationships.\\
Social network connectivity provides information about a user’s relationships and the user’s interactions with a network. Social networking data can represent information about a user that can add nuance to the task of this thesis, especially if the user’s relations have also interacted with the website in the past.\\
Users' social media relationships or posts have previously been shown effective at predicting demographics of users such as age \citep{Perozzi2015} and gender \citep{Burger2011}, and with the concatenated models we experienced how the ability to infer properties of users is an important element towards expanding the click-based models. But while one advantage of social media data is the large quantity of data generated by users, not all of that data will be useful for any particular predicting task leading to a large degree of noisy data. 
\pdfbookmark[1]{Persuasive Recommendations}{persuasive}
\subsection{Persuasive Recommendations}\label{subsec:persuasive}
In physical stores the sales or service associate has the opportunity to deliver a personal sales experience by having a dialog with the customer. This is not naturally built into online stores. Adapting the digital sales process to the needs and problems of the individual customer helps both to increase sales and provide a better customer experience. A key element to implement this level of human engagement in an e-commerce setting is to successfully understand a user's intent.\\
Future work can investigate how to customize persuasive messages for individual customers as they navigate the website, focusing on providing every customer a journey that aligns with their needs and intentions.
Here, A/B testing is a useful tool to learn which persuasion strategy works for a customer segments, i.e. experiment with different persuasion messages and observe the responses from the part of the segment receiving A versus B. 
\pdfbookmark[1]{Ethics}{ethics}
\subsection{Ethics}\label{subsec:ethics}
User behaviour modelling brings forward some concerns about ethics.\\
Log data collected for user modelling purposes is personal information and thus affects privacy. By law, we have to account for user privacy, because privacy is a fundamental human right.
Collection and storage of log data must be done in a legally compliant way adhering to EU and Danish regulation on data handling. 
User behaviour modelling involves the processing of personal data. Processing of personal data is permitted on the basis of consent from the person involved and on the basis of the justified interest of the accountable company. The person involved must always be informed, and the person involved always has the right to appeal and must also be explicitly informed of this right. But even though a company is legally obliged to be transparent in their collection and use of log data, and user behaviour modelling is advantageous for providing quality recommendations, it raises a threat to user privacy. Therefore, an ethical and trustworthy company should also aim to develop a system that provides both high quality recommendations and preserves users' privacy.
There exists some research to achieve the goal of preserving user privacy while
collecting personal information. Differential privacy, introduced by \cite{Dwork2006}, provides a mathematical process for adding randomness to statistical queries with a quantifiable degree of privacy for individuals joining a database. Another approach in privacy preservation of personal data is $k$-anonymity \citep{Sweeney2002}. A collection of data is said to have the $k$-anonymity property if the information for each person contained in the collection cannot be distinguished from at least $k-1$ individuals whose information also appear in the collection.


\medskip

\bibliography{references}

\clearpage

\end{document}